\documentclass[10pt,journal,compsoc]{IEEEtran}

\usepackage{cite}
\usepackage{amsmath}
\usepackage{amsfonts}
\usepackage{graphicx,epsfig,epstopdf}
\usepackage{enumerate}
\usepackage{cuted}
\usepackage{multirow}
\usepackage{multicol}
\usepackage{lipsum}
\usepackage{widetext}
\usepackage{subfigure}
\usepackage{color}

\def\bh{{\mathbb{H}}}
\def\bw{{\mathbb{W}}}
\def\bf{{\mathbb{F}}}

\def\pd{{p_{\mathbf{d}}}}

\def\IEEEraisesectionheading#1{\noindent\raisebox{1.5\baselineskip}[0pt][0pt]{\parbox[b]{\columnwidth}{#1\unskip\global\everypar=\everypar}}\vspace{-1\baselineskip}\vspace{-\parskip}\par}
\def\bbe{{{\mathbb{E}}}}
\newcommand{\dfs}{d_{fs}}
\newcommand{\nft}{n_{ft}}
\newcommand{\dws}{d_{ws}}
\newcommand{\nwt}{n_{wt}}
\newcommand{\hfs}{h_{fs}}
\newcommand{\hfsx}{h_{fs}(x)}

\newcommand{\hft}{h_{ft}}
\newcommand{\hftx}{h_{ft}(x)}

\newcommand{\hws}{h_{ws}}
\newcommand{\hwsx}{h_{ws}(x)}

\newcommand{\hwt}{h_{wt}}
\newcommand{\hwtx}{h_{wt}(x)}

\allowdisplaybreaks
\begin{document}


\title{Information Propagation in \\ Clustered Multilayer Networks}

\author{ \IEEEauthorblockN{Yong Zhuang and Osman Ya\u{g}an} \\\IEEEauthorblockA{Department of ECE and CyLab, Carnegie Mellon University, Pittsburgh, PA 15213 USA \\\{yongzhua, oyagan\}@andrew.cmu.edu}}

%
%

\markboth{IEEE TRANSACTIONS ON NETWORK SCIENCE AND ENGINEERING, 2015}%
{Shell \MakeLowercase{\textit{et al.}}: Bare Demo of IEEEtran.cls for Computer Society Journals}
%



\IEEEtitleabstractindextext{%
\begin{abstract}
In today's world, individuals interact with each other in more complicated patterns than ever.
Some individuals engage through online social networks (e.g., Facebook, Twitter), while some communicate only through conventional ways (e.g., face-to-face).
Therefore, understanding the dynamics of information propagation among humans calls for a multi-layer network model where an online social network is conjoined with  a physical network. 
In this work, we initiate a study of information diffusion in a {\em clustered} multi-layer network model, where all constituent layers are random networks with high clustering.
We assume that information propagates according to the SIR model and with different information transmissibility across the networks.
We give results for the conditions, probability, and size of information epidemics, i.e., cases where information starts from a single individual and reaches a positive fraction of the population. We show that increasing the level of clustering in either one of the layers increases the epidemic threshold and decreases the final epidemic size in the whole system. An interesting finding is that information with low transmissibility spreads more effectively with a small but densely connected social network, whereas highly transmissible information spreads better with the help of a large but loosely connected social network. 


\end{abstract}

\begin{IEEEkeywords}
Information Propagation, Clustered Multilayer Networks, Percolation Theory, Random Graphs.
\end{IEEEkeywords}
}

\maketitle

\IEEEdisplaynontitleabstractindextext
\IEEEpeerreviewmaketitle

\IEEEraisesectionheading{\section{Introduction}\label{sec:introduction}}

The study of dynamical processes on real-world complex networks has been an
active research area over the past decade.  An interesting 
phenomenon that occurs in many such processes is the spreading of 
an initially localized effect throughout the whole (or, a very large part of the) network. 
These events are usually referred to as (information) {\em cascades} and
can be observed in processes as
diverse as adoption of cultural fads, the diffusion of belief, norms, and innovations in social networks 
\cite{WattsExternal,DoddsWatts}, 
disease contagion in human and animal populations \cite{Murray, AndersonMay,SahnehScoglioMieghem,OY13},
failures in {\em interdependent} power systems 
\cite{brummitt2012suppressing,Buldyrev,Vespignani,YaganQianZhangCochranLong},
rise of collective action to joining a riot \cite{Granovetter}, and the global spread of computer viruses 
or worms on the Web \cite{NewmanForrestBalthrop,BalthropForrestNewmanWilliamson}.

This work focuses on an important class of dynamical process known as the information propagation or {\em simple} contagions; this is to be contrasted
with {\em complex} contagions often referred to as {\em influence} propagation \cite{YaganPRE}.
Although well-studied in the past across various domains, the information diffusion problem has recently taken a new form and dimension by the emergence of online social networks such as 
Facebook, Twitter, etc. In particular, due to the existence of multiple online social networks, information is now likely to spread among the population in an unprecedented speed and scale. 
Although there has been a recent surge of research on multi-layer and multiplex networks (e.g., see
\cite{kivela2014multilayer,SahnehScoglioMieghem,OY13}), the current literature still falls short in fully quantifying this phenomenon.  
For instance, Ya\u{g}an et al. analyzed \cite{OY13,yaugan2012information} the diffusion of information in a multi-layer network, but only for the cases where all constituent layers 
are generated by  the {\em configuration model} \cite{MN01}; see also \cite{LeichtDSouza,MendiolaSerranoBoguna} 
for works that are in the same vein. 
However, the configuration model produces \cite{JM09, MN09} networks that can not accurately capture some important aspects of real-world social networks, most notably the property of {\em high
clustering} \cite{WattsStrogatz,SerranoBoguna2}. Informally known as the phenomenon that 
\lq\lq friends of our friends" are likely to be our friends, clustering has been shown to impact significantly 
the dynamics of various diffusion processes \cite{hackett2011cascades,YaganPRE,huang2013robustness}.


 
With these in mind, we study information propagation in {\em clustered} multi-layer networks.
In particular, we consider a model where all constituent layers are {\em random networks with clustering}
as introduced by Miller \cite{JM09} and Newman \cite{MN09}, i.e., they are generated {\em randomly} from 
given distributions specifying the number of single edges and triangles for any given node; see Section \ref{sec:networkClustering}
for details. Our modeling framework consists of 
a physical network 
where information spreads amongst people
through {\em conventional} communication media (e.g.,
face-to-face communication, phone calls),
 and {\em overlaying}
this network, there are online social
networks offering alternative platforms for information diffusion, such 
as Facebook, Twitter, Google+, etc. 
The coupling across these networks results from nodes they have in common, i.e., individuals who participate in multiple networks simultaneously;
see Section \ref{sec:multilayerNetworkModelsWithClustering} for details of our multi-layer network model where the coupling level is {\em tunable}.

In this setting, we analyze the propagation of information assuming that 
information propagates according to the SIR epidemic model\footnote{The analogy between the spread of diseases
and information has long been recognized \cite{DG04} and
the SIR epidemic model is commonly used in similar studies; e.g., see \cite{KL06, XS12}.}.
Namely, an
individual is either {\em susceptible} (S) meaning that she has not
yet received a particular information, or {\em infectious} (I) meaning
that she is aware of the information and is spreading it to
her contacts, or {\em recovered} (R) meaning that she is no
longer spreading the information. 
Let $T_{ij}$ denote the probability that an infectious individual $i$ transmits the
information to a susceptible {\em contact} $j$.
Throughout, we account for the fact that individuals'  
information spreading behaviors may differ from one network to another; e.g., one may be more active in Facebook than Twitter, or vice versa.  
The varying rate of information diffusion across different social networks is captured in our formulation by 
having the transmissibility $T_{ij}$ depend on the network that the link $i\sim j$ belongs to; see Section \ref{sec:infoPropModel} for details.

Our main contributions are as follows. We solve analytically for the threshold, probability, and mean size of {\em information epidemics}, i.e., cases where information starts from a single individual and reaches a positive fraction of the population; see Section \ref{sec:problem} for precise definitions. Our analytical approach is based on
mapping the SIR propagation model to a {\em bond percolation} process and then utilizing a multi-type branching process to solve for the quantities of interest;
the isomorphism between the SIR model and bond percolation has been established for certain cases in \cite{MillerProb,KenahRobins}. The analytical results are validated and extended by computer simulations. 

Several interesting conclusions are drawn from these results. For example, we show
that increasing the level of clustering in any one of the layers increases the epidemic threshold and decreases the final epidemic size of the whole system.
Put differently, we show that i) clustering makes it more difficult for a single person to spread the information to the {\em masses}; and ii) even if the information reaches to the masses, we show that clustering decreases the total fraction of individuals informed. We also demonstrate how the overlap between the constituent networks affect the information propagation dynamics, particularly through impacting the degree-degree correlations. For instance, we show that an online social network that is small in size but large in mean connectivity is more effective (resp.~less effective) in facilitating the propagation of information with {\em low transmissibility} (resp.~{\em high transmissibility}) as compared to a large social network with smaller mean connectivity, with the total number of edges fixed in both cases.



Our general framework contains non-clustered multi-layer networks and single-layer clustered networks as special cases. In addition, given that information propagation problem is studied via bond percolation over a multi-layer network, our work can also be useful in the context of robustness against {\em random attacks}.
Finally, although the problem is motivated here in the context of information propagation, network coupling is relevant in many {\em simple contagion} processes including diffusion of diseases \cite{SahnehScoglioMieghem,LeichtDSouza}; e.g., a small community may consist of three coupled networks corresponding to three venues people can interact at:  households, hospitals, and schools \cite{bansal2006comparative,allard2012bond}.

The rest of the paper is organized as follows. In Section \ref{sec:problemsAndModel}, we introduce the models applied in this study and the problem to be considered.
In Section \ref{sec:relatedTechnologiesAndWork}, we introduce the related technical background.
In Section \ref{sec:mainResults}, we present and derive the main results of this work, while
in Section \ref{sec:numericalResults}, we confirm our analytical results via computer simulations.
We conclude the paper in Section \ref{sec:conclusion}.
 
\section{Problem Formulation}
\label{sec:problemsAndModel}
In this section, we give precise definitions of our system model and then describe problems that shall be studied. 


\subsection{Random Graphs with Clustering}
\label{sec:networkClustering}

\begin{figure}[h]
    \centering
    \includegraphics[width=0.45\textwidth]{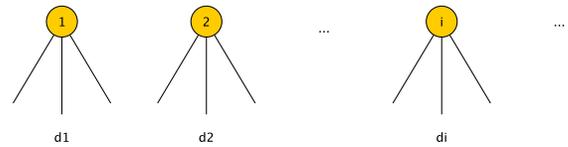}
    \caption{Illustration of process of Configuration Model.
    }
\label{fig:configurationModel}
\end{figure}

Our modeling framework is based on {\em random} networks with {\em clustering} as introduced independently by Miller \cite{MM95}
and Newman \cite{MN03}. This model takes its roots from the widely used 
{\em configuration model} \cite{MN01} that generates a network randomly according
to a given degree distribution. 
Namely,
consider a vertex set $V = {1, 2, \dots, n}$, where each vertex 
is independently assigned a random number of {\em stubs} according to a probability distribution $\{p_k\}_{k=0}^{\infty}$; i.e., the degree $d_i$ of vertex $i$ equals $k$ with probability $p_k$ for any positive integer $k$. 
Then, stubs are randomly paired with each other to form edges until no free stubs is left; see Figure \ref{fig:configurationModel} for an illustration of the configuration model.

It is known that \cite{MN01,MN03} configuration model generates tree-like graphs with number of cycles approaching to zero as the number of nodes gets large. However, most social networks exhibit {\em high clustering}, anecdotally known as the likelihood of a \lq\lq friend of a friend" to be one's friend. Put differently, real-world social networks are not tree-like and instead have considerable number of cycles, particularly of size three; i.e., {\em triangles}.
With this in mind, Miller \cite{MM95}
and Newman \cite{MN03} proposed a modification on the configuration model to enable generating random graphs with given degree distributions and tunable clustering.

\begin{figure}[h]
    \centering
    \subfigure[]{\hspace{-0.5cm} \includegraphics[totalheight=0.12\textheight,width=0.15\textwidth] {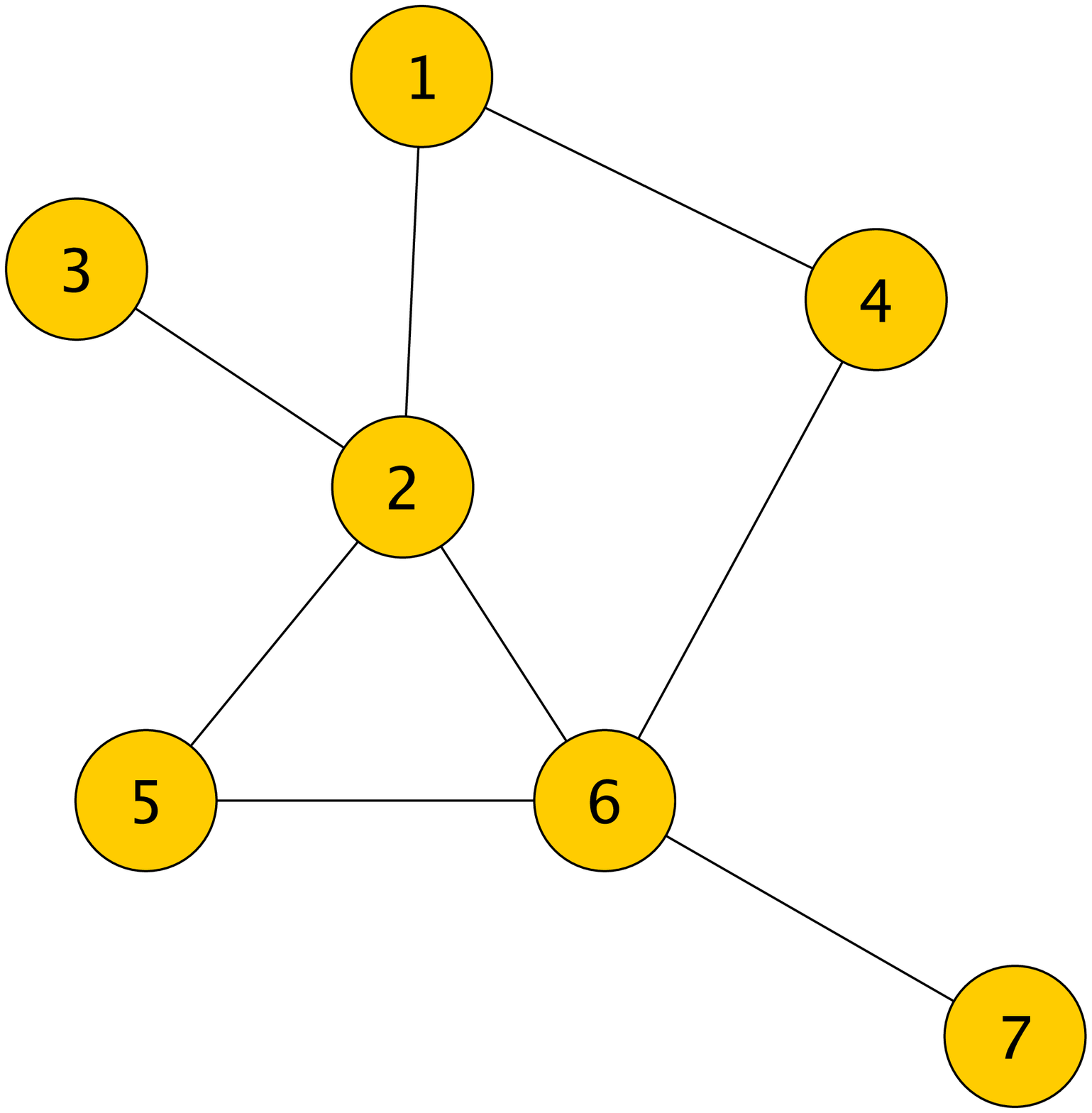}}
    \label{fig:clusteringA}
    \subfigure[]{\hspace{2cm} \includegraphics[totalheight=0.12\textheight,width=0.15\textwidth] {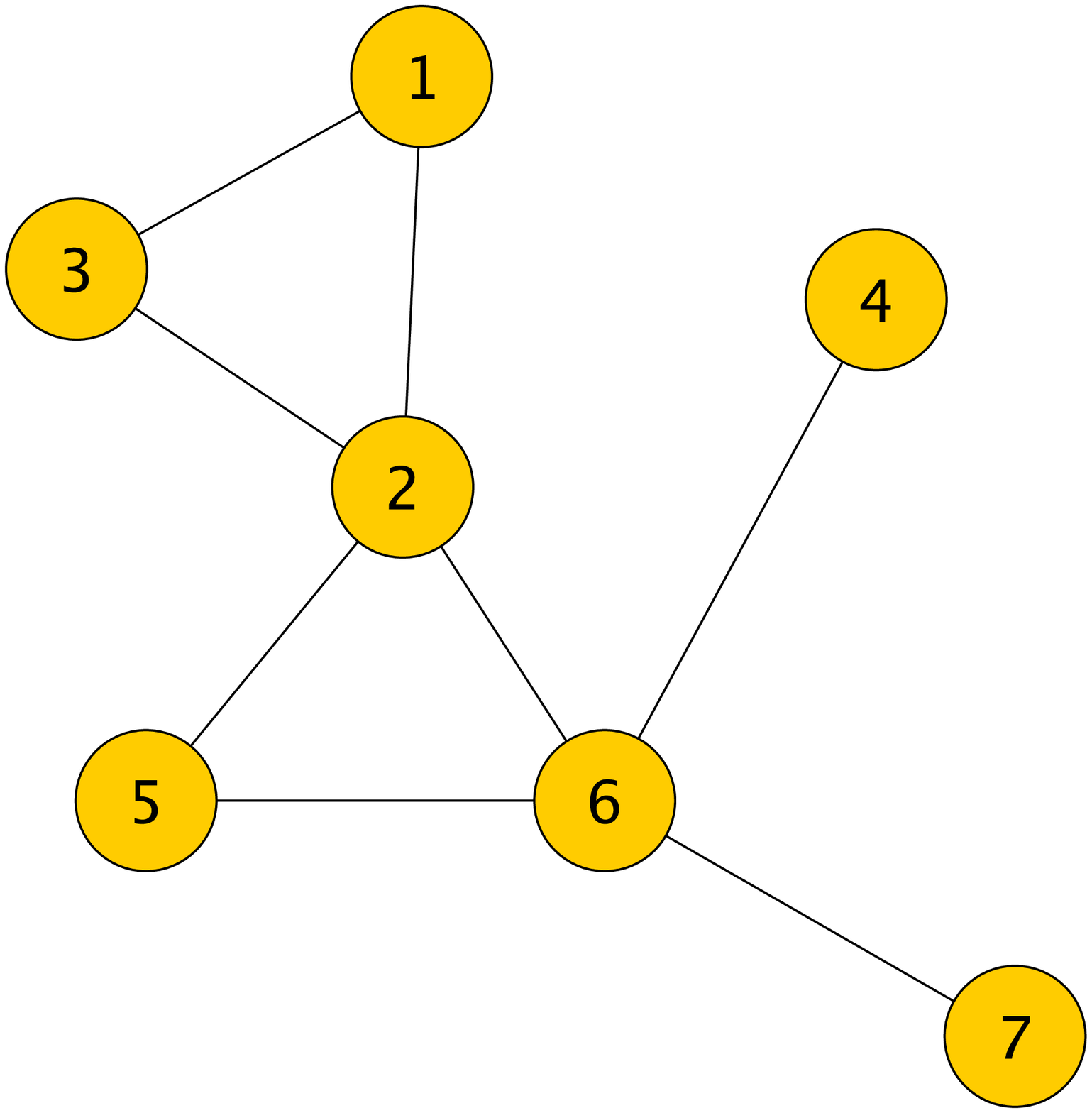}}
    \label{fig:clusteringB}
    \caption{ Illustration of a random network with clustering. In part
  (a), node $2$ has 2 single edges and one triangle, whereas in part (b) it has
  zero single edges and two triangles. 
 For the network in Figure \ref{fig:clusteringA}, the global clustering coefficient is 0.2 while the local clustering coefficient is 0.3, while for the network in  Figure \ref{fig:clusteringB}, these coefficients are given by 0.4 and 0.7, respectively. 
    }
\label{fig:clusteringC}
\end{figure}

The model proposed in \cite{MM95, MN03} is often referred to as random networks with clustering and is
based on the following algorithm.
Consider a joint degree distribution $\{p_{st}\}_{s,t=0}^{\infty}$ that gives
the probability that a node has $s$ single edges and $t$ triangles; e.g., see node $2$ in Figure \ref{fig:clusteringA} that has two single edges and one triangle. Namely, each node will be given $s$ stubs labeled as {\em single} and $2t$ stubs labeled as {\em triangles} with probability $p_{st}$, for any $s,t=1,2,\ldots$. Then, stubs that are labeled as single are {\em randomly} joined to form single edges that are not part of a triangle, whereas {\em pairs} of triangle stubs from three nodes are {\em randomly} matched to form triangles between the three participating nodes; of course the total degree of a node will be distributed by $p_k = \sum_{s,t: s+2t = k} p_{st}$.
As in the standard configuration model, it can be shown that the number of cycles formed by single edges goes to zero as $n$ gets large, and so does the number of cycles of length larger than three \cite{MN01}. 

The resulting level of clustering of the model described above can be quantified in a number of ways. 
Here we consider two widely used metrics known as the  \textit{global} clustering coefficient \cite{MN03} 
and \textit{local} clustering coefficient \cite{AB00};
see \cite{BJ04,JG09,AS05} for other definitions of clustering coefficient proposed in the literature.
Namely, the global clustering is defined via
\begin{align}
    C_{\textrm{global}} = \frac{3 \times (\text{number of triangles in network})}{\text{number of connected triples}},
\end{align}
where \lq\lq connected triples'' means a single vertex connected by edges to two others. 
On the other hand, the local clustering is defined as the average
\begin{align}
    C_{\textrm{local}} = \frac{1}{n^{*}}\sum_{i}C_{i},
\end{align}
where $C_i$ denotes the clustering coefficient for node $i$ given by
\begin{align}
    &C_{i} = \frac{\text{number of triangles connected to vertex }i}{\text{number of connected triples centered on vertex }i}.
    \label{eq:local_clustering_defn}
\end{align}
Here, $n^{*}$ is the number of nodes whose $C_{i}$ is well-defined in the network; i.e., number of nodes where the denominator at (\ref{eq:local_clustering_defn}) is nonzero. The difference between the two definitions of clustering
is illustrated in Figures \ref{fig:clusteringA} and \ref{fig:clusteringB}, where networks with the same degree distribution 
are considered.

It was shown in \cite{MN01} that both  $C_{\textrm{global}}$ and $C_{\textrm{local}}$ are positive in the random clustered network model, while both quantities approach to zero with increasing network size in the standard configuration model. 


\begin{figure*}[!t]
    \centering
    \includegraphics[width=0.65\textwidth]{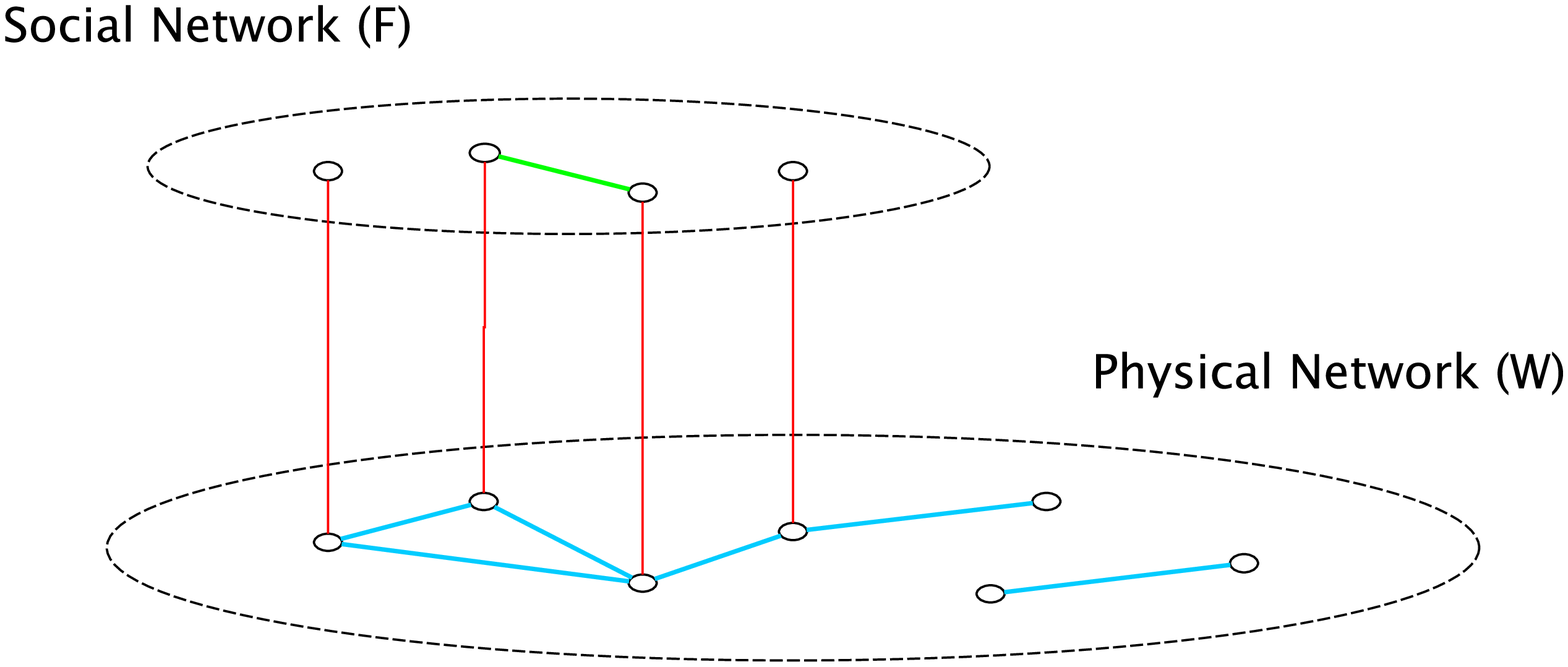}
    \caption{Nodes in the upper circle and lower circle indicate the individuals in \textit{social network} and \textit{physical network} respectively.
    The nodes connected by a red line cross two networks mean they are the same individual existing in two networks.
   Green nodes in the upper circle belong to $\bf$, while blue nodes in $\bw$.
Some of the nodes connected across the two networks by a red line indicates the fact that they represent the same individual. }
    \label{fig:multilayerNetwork}
\end{figure*}

\subsection{Multilayer Network Models with Clustering}
\label{sec:multilayerNetworkModelsWithClustering}
In this paper, we consider a multilayer network where each layer is generated independently and constitutes a random graph with clustering as introduced in Section \ref{sec:networkClustering}.
For brevity, we only consider two layers but most of the arguments can easily be extended higher number of layers. 
Namely, we let $\bw$ and $\bf$ denote the two constituent layers of networks with the possible motivation that $\bw$ models the \emph{physical} contact network among individuals, i.e., models face-to-face relationships, while network
$\bf$ stands for an online social network, say Facebook. In line with this terminology, we assume that the network 
$\bw$ is defined on the vertices $\mathcal{N} = \{1, \dots, n\}$,  while $\bf$ contains only a {\em subset} of the nodes in $\mathcal{N}$ to account for the fact that not every individual participates in online social networks; see
Figure \ref{fig:multilayerNetwork} for an illustration of the two-layer network model we are considering.

To specify this model further, we assume that each vertex in $\mathcal{N}$ participates in $\bf$  independently with probability $\alpha \in (0, 1]$, leading by the Strong Law of Large Numbers to
\begin{align}
   \frac{|\mathcal{N}_{\bf}|}{n} \to_{\textrm{a.s.}} \alpha 
\end{align}
where $\mathcal{N}_{\bf}$ denotes the set of vertices in network $\bf$; here $\to_{\textrm{a.s.}}$ denotes convergence
in almost sure sense with $n$ growing unboundedly large. In words, 
this implies that  the fraction of nodes that belong to $\bf$ is $\alpha$ in the large $n$ limit. The case where $|\mathcal{N}_{\bf}| = o(n)$ has been considered in \cite{OY13} and it was shown that most properties pertaining to the propagation information are unaffected by the existence of the upper layer $\bf$; i.e., when the online social network has a negligible size compared to the whole population, it does not impact the threshold or size of information epidemics.

As mentioned already, we assume that both $\bf$ and $\bw$ are random networks with clustering. In particular, we let $\{p_{st}^{f}, s, t = 0, 1, \dots\}$ and $\{p_{st}^{w}, s, t = 0, 1, \dots\}$ denote the joint distributions for single edges and triangles for $\bf$ and $\bw$, respectively. Then both networks are generated independently according to the algorithm described in Section \ref{sec:networkClustering}, and they are denoted respectively by
$\bf = \bf(n;\alpha, p_{st}^{f})$ and $\bw = \bw(n;p_{st}^{w})$.
We define the multi-layer network $\bh$ as the {\em disjoint} union 
$\bh = \bf \coprod \bw$ and represent it by $\bh(n;\alpha, p_{st}^{f}, p_{st}^{w})$. Here, the disjoint union operation implies that we still distinguish $\bf$-edges from $\bw$-edges in network $\bh$, and this is done to accommodate the
possibly different rates (or, even rules) of information propagation across the two networks. To this end, an equivalent representation of $\bh$ would be a {\em multiplex} network with different types (or, colors) of edges.

With these definitions in mind, let $\dfs$ and $\dws$ to denote the random variables corresponding to the number of \textit{single edges} for a vertex in $\bf$ and $\bw$, respectively, while $\nft$ and $\nwt$ are defined similarly for the number of triangles assigned; i.e., the degree of a node from triangle {\em edges} in $\bf$ is given by $d_{ft}=2n_{ft}$ and similarly for $d_{wt}$.
Then the {\em colored} degree $\mathbf{d}$ of a vertex is given by
\begin{align}
\mathbf{d} = (d_{fs}, 2n_{ft}, d_{ws}, 2n_{wt})
\end{align}
meaning that the vertex has $d_{fs}$ \textit{single edges} and $2n_{ft}$ \textit{triangle edges} in network $\bf$, and $d_{ws}$ \textit{single edges} and $2n_{wt}$ \textit{triangle edges} in network $\bw$.
Under the assumptions enforced here, the distribution of this colored degree is given by
\begin{align}
    p_{\mathbf{d}} = \left(\alpha p^f_{d_{fs}n_{ft}} + (1 - \alpha) \mathbf{1}[d_{fs} = 0 ~\wedge~ n_{ft} = 0]\right)p^w_{d_{ws}n_{wt}}
    \label{eq:possibilityPd}
\end{align}
where the term $ (1 - \alpha) \mathbf{1}[d_{fs} = 0 ~\wedge~ n_{ft} = 0]$ accounts for the fact that if the node
does not belong to $\bf$ (which happens with probability $1-\alpha$), then its degree from single and triangle edges will both be zero.

\subsection{Information Propagation Model: SIR}
\label{sec:infoPropModel}

Consider the diffusion of a piece of information in the multi-layer network $\mathbb{H}$
which starts from a single node.
We assume that information spreads from a node to its neighbors according
to the SIR epidemic model. In this context, an
individual is either {\em susceptible} (S) meaning that she has not
yet received a particular item of information, or {\em infectious} (I) meaning
that she is aware of the information and is capable of spreading it to
her contacts, or {\em recovered} (R) meaning that she is no
longer spreading the information \cite{DG04,KL06, XS12}.
As in
\cite{MN02},
we assume that an infectious individual $i$ transmits the
information to a susceptible contact $j$ with probability $T_{ij} = 1-e^{-r_{ij}\tau_i}$.
Here, $r_{ij}$ denotes the rate of contact over
the link from $i$ to $j$, and $\tau_i$ is the time $i$ keeps
spreading the information; i.e., time $i$ remains
infectious.

It is expected that  information propagates over the physical
and social networks at different rates, which manifests from
different probabilities $T_{ij}$ across links in this case.
Specifically, let $T^w_{ij}$ stand for the probability of
information transmission over a link (between and $i$ and $j$) in
$\mathbb{W}$ and let $T^f_{ij}$ denote the probability of
information transmission over a link in $\mathbb{F}$. For
simplicity, we assume that $T^w_{ij}$ and $T^f_{ij}$ are
independent for all distinct pairs $i,j=1, \ldots, n$.
Furthermore, we assume that the random variables $r^{w}_{ij}$ and
$\tau^{w}_{i}$ are independent and identically distributed
(i.i.d.) with probability densities $P_w(r)$ and $P_w(\tau)$,
respectively. We find it useful to define $T_w$ as the
mean of $T^{w}_{ij}$; i.e.,
\[
T_w:=  \langle T^{w}_{ij} \rangle  = 1 - \int_{0}^{\infty}\int_{0}^{\infty} e^{-r
\tau}P_w(r)P_w(\tau) dr d\tau.
\]
We refer to $T_w$ as the {\em transmissibility} 
over  $\mathbb{W}$ and note that $0 \leq T_w \leq 1$. 
In the same manner, we assume that $r^{f}_{ij}$ and
$\tau^{f}_{i}$ are i.i.d. with respective densities $P_f(r)$ and
$P_f(\tau)$ leading to a transmissibility $T_f$ over $\mathbb{F}$.

As shall be discussed in Section \ref{sec:problem}, under certain conditions,
it can be assumed that information propagates over $\mathbb{W}$ (resp.~over $\mathbb{F}$) as if all
transmission probabilities were equal to $T_w$ (resp.~to $T_f$), for the purposes of computing the threshold, probability, and expected size of  epidemics. 
 
\subsection{Problems of Interest}
\label{sec:problem}
We consider the propagation of information (or, a disease) in $\mathbb{H}$ as explained in Section \ref{sec:infoPropModel}.
The outbreak is triggered by infecting a randomly selected node and propagates in the network according to the SIR model. Given the monotonicity of the SIR process \cite{MN02}, a steady-state will always
 be reached where all nodes are either {\em recovered} or {\em susceptible}.
The final size of an outbreak is defined as the number of nodes that are {\em recovered} at the steady-state, and its {\em relative} final size is its final size divided by the total size $n$ of the network.  Following \cite{KenahRobins}, we define a {\em self-limited outbreak} as an outbreak whose relative final size approaches zero, and an {\em epidemic} to be an outbreak whose relative final size is positive, both in the limit of large $n$ . There is a {\em critical boundary} in the space of all network parameters, often defined as the 
epidemic {\em threshold}, or epidemic boundary, that separates the cases for which the probability of an epidemic is zero (i.e., {\em sub-critical}, or non-epidemic parameter regime)
from those that lead to $\mathbb{P}[\textrm{epidemic}] >0$ (i.e., {\em super-critical}, or epidemic regime), again with $n \to \infty$.

With these definitions in place, this work seeks to identify i) the epidemic boundary; ii) the relative final size of epidemics in the super-critical case; and iii) the exact probability $\mathbb{P}[\textrm{epidemic}]$ in the super-critical regime. 

As we seek to study several properties of simple contagions as outlined above, a first step will be to observe that 
under certain conditions, the SIR propagation model is {\em isomorphic} to a {\em bond percolation} process \cite{BroadbentHammersley}.
More specifically, assume that each edge 
in $\mathbb{W}$ (resp. $\mathbb{F}$) is {\em
occupied} -- meaning that it can be used in spreading the
information, disease, etc. -- with probability $T_w$ (resp. $T_f$) independently from all other
edges. Here, $T_w$ and $T_f$ are transmissibility parameters calculated as the mean 
probability of transmission between any two nodes in the corresponding networks; see Section \ref{sec:infoPropModel}.
Then, the size of an outbreak started from an arbitrary node 
is equal to the number of
individuals that can be reached from the initial node by using
only the {\em occupied} links in $\mathbb{H}$. 

This isomorphism was claimed to hold first by Newman \cite{MN02} who studied the SIR model in single networks.
It was later shown by several authors \cite{MillerProb,KenahRobins} that the SIR process is isomorphic to bond percolation
{\em only} when the infectious period distribution $P(\tau)$ is {\em degenerate}; i.e., when  all nodes have the same recovery time $\tau_1=\cdots = \tau_n$. 
When nodes have heterogeneous recovery times, the SIR process is {\em not} isomorphic to a bond percolation process. However, \cite{KenahRobins,MillerProb} proved that, in the large 
network size limit, a bond percolation process can still be used to accurately predict 
the a) epidemic boundary, b) mean size of self-limited out-breaks, and c) relative final size of epidemics.
With respect to our goals, it is only the probability $\mathbb{P}[\textrm{epidemic}]$ that can't be obtained through analyzing the bond percolation model when the recovery times are heterogeneous; in fact, we are not aware of any technique in the literature that enables calculating $\mathbb{P}[\textrm{epidemic}]$ exactly in these cases. 
Therefore, we restrict our attention to cases where the recovery times are uniform when dealing with $\mathbb{P}[\textrm{epidemic}]$, while more general cases are considered for the boundary and final size of epidemics.
To that end, our efforts towards analyzing information propagation (e.g., items (i)-(iii) given above) rely on mapping the SIR model to a bond percolation process.

%

We now explain how mapping the problem to a bond percolation process paves the way to obtaining the quantities (i)-(iii) given above. 
Let $\mathbb{\tilde{W}}$ (resp.~$\mathbb{\tilde{F}}$) be a network that contains only the occupied edges of $\mathbb{W}$
(resp.~$\mathbb{F}$). Put differently, consider an Erd\H{o}s-R\'enyi \cite{Bollobas}
network $\mathbb{G}(n;T_w)$ (resp.~$\mathbb{G}(\mathcal{N}_F;T_f)$) on the nodes $\{1,\ldots, n\}$ (resp.~on the node set 
$\mathcal{N}_F$) such that between every pair of nodes there is an edge
with probability $T_w$ (resp.~$T_f$) 
independently from all other edges. Then,
$\mathbb{\tilde{W}} = \mathbb{W} \cap (\mathbb{G}(n;T_w))$ and $\mathbb{\tilde{F}} = \mathbb{F} \cap (\mathbb{G}(\mathcal{N}_F;T_f))$.
The bond percolation network $\mathbb{\tilde{H}}$ that contains only the occupied edges of $\mathbb{H}$ is then given by 
$\mathbb{\tilde{H}} = \mathbb{\tilde{W}} \cup \mathbb{\tilde{F}}$. 
The different transmissibility properties of
$\mathbb{W}$ and $\mathbb{F}$ are already incorporated into this model through distinct bond occupation probabilities $T_w$ and $T_f$.
Thus, $\mathbb{\tilde{H}}$ (defined on the vertices $\{1,\ldots, n\}$) 
is a {\em simplex} network obtained by a simple union of the edges of $\mathbb{\tilde{W}}$ and
$\mathbb{\tilde{F}}$.

The threshold and relative final size of epidemics can now be computed
from the {\em phase transition} behavior of $\mathbb{\tilde{H}}$.
Namely, epidemics can take place if and only if 
$\mathbb{\tilde{H}}$ has a {\em giant} component; i.e., a {\em connected} subgraph
that contains a {\em positive} fraction of nodes in the large $n$ limit. Thus, epidemic boundary 
is given by the phase transition threshold, i.e., the threshold for the existence of a giant component in $\mathbb{\tilde{H}}$.
Also, a node
can trigger an epidemic only if it belongs to the giant component, in which case 
an outbreak started from this node will reach the whole giant component.
Hence, the relative size of the giant component in
$\mathbb{\tilde{H}}$ gives {\em both} $\mathbb{P}[\textrm{epidemic}]$ as well as the relative final
size of epidemics. 

\section{Technical Background}
\label{sec:relatedTechnologiesAndWork}

\begin{figure}[!h]
    \centering
    \includegraphics[width=0.42\textwidth]{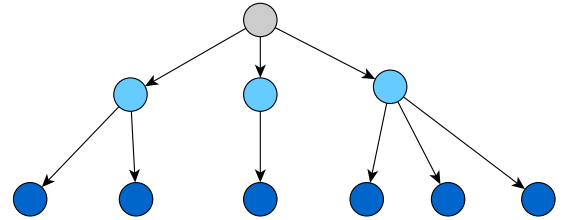}
        \caption{Illustration of the branching process.
        The children of each individual node is identified recursively, while taking into 
        account whether or not the information is transferred from the parent node to the child node. The initial vertex that starts the information is regarded as the $0^\text{th}$ generation, and we are interested in deriving the limiting behavior of the total number of nodes reached and informed as the number nodes $n \to \infty$.
        }
    \label{fig:branchingProcess}
\end{figure}

In what follows we introduce the technical underpinnings of our analysis. Our approach is based on exploring a branching process which starts with an arbitrary node in the network and recursively reveals all the nodes reached {\em and} informed by following its edges; see Figure \ref{fig:branchingProcess}. Throughout, we will be interested in various discrete random variables naturally associated with this branching process; e.g., total number of nodes reached and informed by following a randomly selected edge in $\mathbb{W}$ (resp.~in $\mathbb{F}$).
Oftentimes we find it useful to characterize the probability distributions of these random variables through their {\em generating functions} \cite{HW13}. This approach has been widely adopted in the literature in analyzing complex networks and has several benefits as shall soon become apparent.  




We now formally define the notion of a generating function: Let $X$ be a positive-valued, discrete random  variable with the distribution 
$\{p_{k}: k=0,1,\ldots \}$; i.e., we have $\mathbb{P}(X=k)=p_k$. 
Then the generating function of $X$ is given by
\begin{align}
    h(x) = \sum_{k = 0}^{\infty} p_{k} x^{k}, \quad x \in \mathbb{R}.
    \label{func:generatingFunctions}
\end{align}
We remark that a random variable is uniquely identified by its generating function since we have
\[
p_k = h^{(k)}(x) / k!, \qquad k=0,1,\ldots
\]
where $h^{(k)}(x)$ denotes the $k^{\textrm{th}}$ order derivative of $h(x)$. Also, we can easily compute the moments of $X$ from the derivatives of $h(x)$ evaluated at the point $x=1$ \cite{MN01}; e.g., the first moment is given through $\mathbb{E}[X] = h'(1)$, i.e., by the first derivative of $h(x)$ evaluated at $x=1$.

%
%
%


\section{Main Results}
\label{sec:mainResults}
As described in Section \ref{sec:multilayerNetworkModelsWithClustering}, the clustered multilayer network in this paper consists of four kinds of edges, \textit{single edges} in  $\bf$, \textit{triangle edges} in $\bf$, \textit{single edges} in $\bw$, and \textit{triangle edges} in $\bw$; these will be denoted by $fs-$, $ft-$, $ws-$, and $wt-$edges, respectively.
In order to analyze the information propagation in multilayer networks, we consider a branching process that
starts with informing a node selected randomly from among all nodes, $\{1,\ldots, n\}$. We then explore all the neighbors that are reached and informed by this node, and continue recursively until the branching process stops. The distribution of the resulting number of nodes
informed will be characterized via its generating function. 

We now explain our approach based on generating functions precisely. Let $H(x)$ denote the generating function for the \lq\lq {\em finite} number of nodes that are reached and informed" by the above branching process. We will derive an expression for $H(x)$ using four other generating functions $\hfsx$, $\hftx$, $\hwsx$, and $\hwtx$, where
$\hfsx$ stands for the \lq\lq finite number of nodes reached and informed by following a randomly selected $fs$-edge," and $\hwsx$ defined similarly for the $ws-$edges. The definitions for $\hftx$ and $\hwtx$ are a bit different in the sense that they correspond to the \lq\lq finite number of nodes reached and informed by following a randomly selected {\em triangle} in $\bf$ (resp.~in $\bw$)" for $\hftx$ (resp.~$\hwtx$). In other words, we consider the whole triangle at once, rather than focusing on its edges separately; see 
see Section \ref{sec:informationPropagationTriangle}. 


With these definitions in place, we now write $H(x)$ in terms of $\hfsx$, $\hftx$, $\hwsx$, and $\hwtx$:
\begin{align}
	H(x) = x\sum_{\mathbf{d}} p_{\mathbf{d}} h_{fs}(x)^{d_{fs}}h_{ft}(x)^{n_{ft}}h_{ws}(x)^{d_{ws}}h_{wt}(x)^{n_{wt}},
    \label{eq:HX}
\end{align}
where $p_{\mathbf{d}}$ denotes the colored degree distribution given by (\ref{eq:possibilityPd}). 
The validity of (\ref{eq:HX}) can be seen as follows. The term
$x$ stands for the node that is selected randomly and given the information to initiate the propagation. 
This node has a degree $\mathbf{d} = (d_{fs}, 2n_{ft}, d_{ws}, 2n_{wt})$ with probability $p_{\mathbf{d}}$.
The number of nodes reached and informed by each of its $d_{fs}$ (resp.~$d_{ws}$) single edges in $\bf$ (resp.~$\bw$) has a generating 
function $\hfsx$ (resp.~$\hwsx$). Similarly, the number of nodes informed by following each of the $n_{ft}$ (resp.~$n_{wt}$) triangles it participates in $\bf$ (resp.~$\bw$) has a generating function $\hftx$ (resp.~$\hwtx$). Combining, we see from the {\em powers property} of generating functions \cite{MN01} that the number of nodes reached and informed in this process when the initial node has  degree $\mathbf{d}$ has a generating function 
$h_{fs}(x)^{d_{fs}}h_{ft}(x)^{n_{ft}}h_{ws}(x)^{d_{ws}}h_{wt}(x)^{n_{wt}}$.
Averaging over all possible degrees $\mathbf{d}$ of the initial node, we get (\ref{eq:HX}).

For (\ref{eq:HX}) to be useful, we shall derive expressions for the generating functions $\hfsx$, $\hftx$, $\hwsx$, and $\hwtx$. As will  become apparent soon, there are no explicit equations defining these functions. Instead, we should seek for {\em recursive} equations defining each generating function in terms of others. Then, fixed points of this recursion will be explored and utilized to determine the threshold and size of information epidemics; i.e., situations where the number of people reached and informed by the original branching process is {\em infinite}. These steps are taken in the next sections where we first focus on deriving $\hfsx$ and $\hwsx$
(Section \ref{sec:informationPropagationSingle}) followed by derivations of $\hftx$ and $\hwtx$ (Section \ref{sec:informationPropagationTriangle}). These arguments are then combined
in Section \ref{sec:fractionOfTheGiantConnectedComponent} to derive the epidemic threshold and final epidemic size. 


\subsection{Information Propagation via Single Edges in Network $\bf$}
\label{sec:informationPropagationSingle}

We start by deriving recursive equations for $\hfsx$ and $\hwsx$, 
by focusing on the number of nodes reached and informed by following one end of a single edge in
$\bf$ and $\bw$, respectively. For instance, for $\hfsx$, we pick one of the single edges in $\bf$ uniformly at random and assume that it is connected at one end a node who is in the {\em infected} state. Then, we compute the generating function for the number of nodes informed by following the other end of the edge. In what follows, we only derive $\hfsx$ since the computation of $\hwsx$ follows in a very similar manner. 

Similar to \cite{OY13}, we obtain the following expression for the generating function $\hfsx$:
\begin{align}
\label{eq:hfsx1}
	& \hfsx 
	\\
	  & = T_{f}x\sum_{\mathbf{d}} \frac{\dfs p_{\mathbf{d}}}{\langle \dfs \rangle} \hfs(x)^{d_{fs} - 1}\hft(x)^{\nft}\hws(x)^{\dws}\hwt(x)^{\nwt} 
	  \nonumber \\
	  & ~~~~ + (1 - T_{f}).
	  \nonumber
\end{align}
We now explain each term appearing at (\ref{eq:hfsx1}) in turn. 
First of all, it is straightforward to see that if the selected edge is {\em not occupied}, which happens with probability $1-T_f$, then the number of informed nodes by following it will be zero.
This leads to a term $(1-T_f)x^{0}$ in the generating function $\hfsx$. In words, adding the term
$(1-T_f)x^{0}$ to $\hfsx$ means that the probability of the underlying random variable (encoded by the generating function $\hfsx$) being zero is incremented by $1-T_f$. On the other hand, if the 
selected edge is occupied, which happens with probability $T_f$, then the node at the other end of the edge will be informed. This means that the number of informed edges in this process will be one {\em plus} all the nodes that are then informed by the node at the other end of the selected edge. Adding one to a random variable is equivalent to multiplying its generating function by $x$, whence we get the term $T_f x$. 

The summation term appearing
at (\ref{eq:hfsx1})
stands for the number of nodes informed by the aforementioned {\em end node} of the randomly selected edge,
and  is similar in vein with the summation term used in (\ref{eq:HX}) with two differences. First, the degree distribution of this end node is not $p_{\mathbf{d}}$ since it is already known to have at least one single edge in $\bf$. Instead, its degree distribution will be proportional to $ d_{fs} p_{\mathbf{d}}$, and after proper normalization we see that the end node will have degree $\mathbf{d} = (d_{fs}, 2n_{ft}, d_{ws}, 2n_{wt})$ with probability 
$\frac{\dfs p_{\mathbf{d}}}{\langle \dfs \rangle}$; e.g., see \cite{OY13,MN01} for similar arguments.
Finally, if this node has degree $\mathbf{d}$ then the number of people it informs is generated by 
$\hfs(x)^{d_{fs} - 1}\hft(x)^{n_{ft}}\hws(x)^{d_{ws}}\hwt(x)^{d_{wt}}$, with the minus one term on $d_{fs}$ accounting to the fact that one of its single edges in $\bf$ has carried the information to this node and has already been taken into account. Averaging over all possible $\mathbf{d}$, we get (\ref{eq:hfsx1}).

\begin{figure}[!t]
    \centering
    \includegraphics[width=0.23\textwidth]{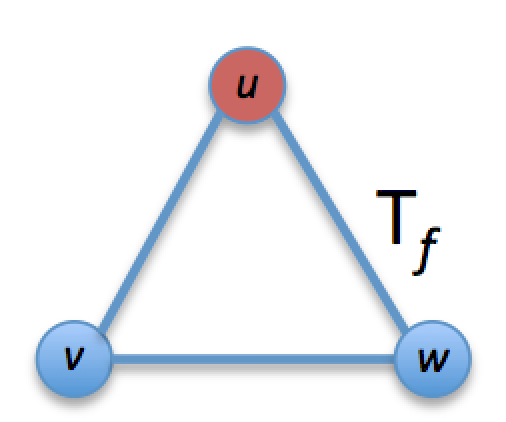}
    \caption{The top vertex $u$ is infected, and information is transferred through an edge only if it is occupied (happens with probability $T_{f}$ in network $\bf$).
    }
    \label{fig:triangleEdge}
\end{figure}

\subsection{Information Propagation via Triangles in Network $\bf$}
\label{sec:informationPropagationTriangle}
We now derive $\hftx$, i.e., the generating function for the number of nodes
informed by following a random triangle in $\bf$; similar arguments hold for $\hwtx$. We demonstrate
this situation in Figure \ref{fig:triangleEdge}, where the top vertex $u$ is {\em infected}, and we are interested in 
computing the generating function for the number of nodes that will be informed {\em by} nodes $v$ {\em and} $w$. 
Firstly, by conditioning on the state, i.e., occupied or not occupied, of the three edges forming this triangle, we compute the probabilities for neither, one, or both of $v$ and $w$ being informed, respectively. It is not difficult to see
that 
\begin{center}
\[
\begin{cases}
  \mathbb{P}[\mbox{none of $v$ and $w$ are informed}] & = (1 - T_{f})^2 \\
   \mathbb{P}[\mbox{one of $v$ and $w$ are informed}] &= 2T_{f}(1 - T_{f})^{2} \\
\mathbb{P}[\mbox{both of $v$ and $w$ are informed}]  &=  2T_{f}^{2}(1 - T_{f}) + T_{f}^{2}.\\
\end{cases}
\]
\end{center}

We now explain why the above equations hold. Firstly, for $v$ and $w$ to be not informed, both of the 
edges $u \sim v$ and $u \sim w$ should be {\em not} occupied. By independence, this occurs with probability
$(1-T_f)^2$. Secondly, we compute the probability of only one of $v$ and $w$ being informed, which by symmetry is given by two times 
the probability that $v$ is informed but $w$ is not. The latter happens if and only if the edge $u \sim v$ is occupied
while the edges $u \sim w$ and $v \sim w$ are not occupied. By independence, this has
probability $T_{f}(1 - T_{f})^{2}$. Finally, probability that both $v$ and $w$ are informed is given by 
subtracting the first two probabilities from one.



We now turn to computing the generating function $\hftx$ by conditioning on the three events discussed above. 
As in Section \ref{sec:informationPropagationSingle}, if neither of the nodes $v$ and $w$ are informed, then the number of nodes informed by this triangle will be zero, leading to an additive term $(1-T_f)^2 x^{0}$. Next, we derive  the term corresponding to the case where only one of $v$ or $w$ is informed. This leads to
        \begin{align}
           & \left(2T_{f}\left(1 - T_{f}\right)\right)x
                      \label{eq:middle_term_triangle}
   \\ 
            & ~\cdot \sum_{\mathbf{d}}\frac{\nft\pd}{\langle n_{ft}\rangle}\hfs(x)^{\dfs}\hft(x)^{\nft - 1}\hws(x)^{\dws}\hwt(x)^{\nwt},
            \nonumber
        \end{align}
where $2T_{f}\left(1 - T_{f}\right)$ stands for the probability of the conditioning event that only one of  $v$ or $w$ is informed, and $x$ stands for the node that is informed. As in Section \ref{sec:informationPropagationSingle}, the degree distribution of this informed node will not be given by $\mathbf{p_d}$, but instead will be proportional to 
the number of triangles $n_{ft}$ assigned to it; as before this is due to the fact that the node under consideration is known to have at least one triangle in $\bf$. By normalization, we see that the degree of the node will be
$\mathbf{d} = (d_{fs}, 2n_{ft}, d_{ws}, 2n_{wt})$ with probability 
$\frac{\nft \pd}{\langle n_{ft}\rangle}$. The rest of the expression (\ref{eq:middle_term_triangle}) follows similarly to
(\ref{eq:hfsx1}), where a minus one term is invoked at $\nft$ in order to not double count the triangle $u , v ,w$ that is being considered.
Finally, the term 
corresponding to the case where {\em both} $v$ and $w$ are informed is easily computed as the {\em square} of (\ref{eq:middle_term_triangle}) as we use the powers property 
upon noting that $v$ and $w$ will inform independent sets of nodes under the enforced assumptions.
Collecting, we obtain
\begin{align}
&\hftx
  \label{func:hftx_osy} \\ \nonumber
& =  (1 - T_{f})^{2}+ \left(2T_{f}\left(1 - T_{f}\right)^{2}\right) x  \sum_{\mathbf{d}} \bigg( \frac{n_{ft}p_{\mathbf{d}}}{\langle n_{ft} \rangle} \hfs(x)^{\dfs}
\\ \nonumber
 & ~~ \cdot \hft(x)^{\nft - 1} \hws(x)^{\dws}\hwt(x)^{\nwt}  \bigg) + \left(2T_{f}^{2}(1 - T_{f}) + T_{f}^{2}\right)
    \notag\\
    & ~~ \cdot \left(x\sum_{\mathbf{d}}\frac{n_{ft}p_{\mathbf{d}}}{\langle n_{ft} \rangle} \hfs(x)^{d_{fs}}\hft(x)^{\nft - 1}\hws(x)^{\dws}\hwt(x)^{\nwt}\right)^{2} 
    \nonumber
\end{align}

\subsection{Computing the Final Epidemic Size}
\label{sec:fractionOfTheGiantConnectedComponent}

We are now in a position to write the recursive equations for generating functions $\hfsx$, $\hftx$, $\hwsx$, and $\hwtx$, whose solution will be reported into (\ref{eq:HX}) to get the final epidemic size. Using
(\ref{eq:hfsx1}) and (\ref{func:hftx_osy}) and similar expressions for $\hwsx$ and $\hwtx$, we obtain

\begin{widetext}
\begin{align}
    \hfsx =\ &T_{f}x\sum_{\mathbf{d}} \frac{\dfs p_{\mathbf{d}}}{\langle \dfs \rangle} \hfs(x)^{d_{fs} - 1}\hft(x)^{\nft}\hws(x)^{\dws}\hwt(x)^{\nwt} + (1 - T_{f}),
    \label{func:hfsx}
    \\
    \hftx =\ & \left(2T_{f}\left(1 - T_{f}\right)^{2}\right) x\sum_{\mathbf{d}} \frac{n_{ft}p_{\mathbf{d}}}{\langle n_{ft} \rangle} \hfs(x)^{\dfs}\hft(x)^{\nft - 1}\hws(x)^{\dws}\hwt(x)^{\nwt}  
    \notag\\
    \ &+ \left(2T_{f}^{2}(1 - T_{f}) + T_{f}^{2}\right) \left(x\sum_{\mathbf{d}}\frac{n_{ft}p_{\mathbf{d}}}{\langle n_{ft} \rangle} \hfs(x)^{d_{fs}}\hft(x)^{\nft - 1}\hws(x)^{\dws}\hwt(x)^{\nwt}\right)^{2}  + (1 - T_{f})^{2},
    \label{func:hftx}
    \\
    \hwsx =\ & T_{w}x\sum_{\mathbf{d}} \frac{d_{ws}p_{\mathbf{d}}}{\langle d_{ws} \rangle} \hfs(x)^{\dfs}\hft(x)^{\nft}\hws(x)^{\dws - 1}\hwt(x)^{\nwt} + (1 - T_{w}),
    \label{func:hwsx}
    \\
    \hwtx =\ & \left(2T_{w}\left(1 - T_{w}\right)^{2}\right) x\sum_{\mathbf{d}} \frac{\nwt p_{\mathbf{d}}}{\langle \nwt \rangle} \hfs(x)^{\dfs}\hft(x)^{\nft}\hws(x)^{\dws}\hwt(x)^{\nwt - 1} 
    \notag\\
    \ &+ \left(2T_{w}^{2}(1 - T_{w}) + T_{w}^{2}\right) \left(x\sum_{\mathbf{d}}\frac{\nwt p_{\mathbf{d}}}{\langle \nwt \rangle} \hfs(x)^{\dfs}\hft(x)^{\nft}\hws(x)^{\dws}\hwt(x)^{\nwt - 1}\right)^{2} + (1 - T_{w})^{2}.
    \label{func:hwtx}
\end{align}
\end{widetext}

The desired generating function $H(x)$ for the finite number of nodes informed in the network can now be computed in the following manner. For  any $x$, we solve the recursive relations (\ref{func:hfsx}) - (\ref{func:hwtx}), i.e., find a fixed point of  (\ref{func:hfsx}) - (\ref{func:hwtx}). Then reporting the resulting values of $\hfsx$, $\hftx$, $\hwsx$, and $\hwtx$ into (\ref{eq:HX}), we obtain $H(x)$ for this particular value of $x$. Repeating the same process for any $x$ will lead to a complete characterization of $H(x)$. However, in this work we are only interested in the cases where the number of nodes informed by the process is {\em infinite}. More precisely, we wish to derive i) the conditions for the probability of informing a positive fraction of nodes to be larger than zero
in the large $n$ limit; and (ii) the exact asymptotic fraction of informed individuals when the conditions of part (i) hold.
As explained in Section \ref{sec:problem}, the latter also gives the probability of triggering an epidemic starting with
a random node.

In order to achieve these goals, we take advantage of the \lq\lq conservation of probability" property of generating functions, i.e., the fact that $H(1)=1$ when the number of nodes reached and informed is always {\em finite}. If on the other hand $H(1)<1$, we understand that there is a positive probability $1-H(1)$ for the aforementioned branching process to lead to an {\em infinite} component of informed nodes; i.e., for the branching process to be {\em supercritical}. In this case, $1-H(1)$ stands for the fraction of nodes that are in the {\em giant component} of $\tilde{\mathbb{H}}$. Recalling the discussion in Section \ref{sec:problem}, we know that information propagation will turn into an epidemic if and only if the initiator node is in this giant component. Thus, we conclude that the probability of an epidemic is given by $1-H(1)$ and so is the relative final size of epidemics. 

With these in mind, we now seek for a fixed point of the recursion (\ref{func:hfsx}) - (\ref{func:hwtx}) at the point 
$x=1$. For notational convenience, we define $h_{1} := h_{fs}(1)$, $h_{2} := h_{ft}(1)$, $h_{3} := h_{ws}(1)$, and $h_{4} := h_{wt}(1)$. The recursion (\ref{func:hfsx}) - (\ref{func:hwtx}) then takes the form
\begin{align}
h_i &= g_i(h_1,h_2,h_3,h_4), \quad i=1,2,3,4
\label{eq:recursion_with_new_notation}
\end{align}
where $g_1$, $g_2$, $g_3$, and $g_4$ are functions immediately obtainable from (\ref{func:hfsx}) - (\ref{func:hwtx});
e.g., we have
\begin{align}
& g_1(h_1,h_2,h_3,h_4) 
\nonumber \\ 
& ~ = T_{f}
\sum_{\mathbf{d}} \frac{\dfs p_{\mathbf{d}}}{\langle \dfs \rangle} h_1^{d_{fs} - 1} h_2^{\nft} h_3^{\dws} 
h_4^{\nwt} + (1 - T_{f}).
\nonumber
\end{align}
With this notation, we also have
\begin{align}
H(1) =  \sum_{\mathbf{d}} p_{\mathbf{d}} h_{1}^{d_{fs}}h_{2}^{n_{ft}}h_{3}^{d_{ws}}h_{4}^{n_{wt}}.
\label{eq:H_with_new_notation}
\end{align}

It is easy to check that the recursion (\ref{func:hfsx}) - (\ref{func:hwtx}) exhibits a trivial fixed point $h_{1}= h_{2} = h_{3} = h_{4} = 1$, which leads to $H(1) = 1$, meaning that the branching process is sub-critical and all informed components have \textit{finite} size.
However, the solution $h_{1} = h_{2} = h_{3} = h_{4} = 1$ is stable only when it is an \textit{attractor}; i.e., a {\em stable} fixed point. We check the stability of this solution via {\em linearization} of (\ref{func:hfsx}) - (\ref{func:hwtx}) around
$x=1$, which leads to Jacobian matrix $\mathbf{J}$ whose entries are given by
\[
\mathbf{J}(i,j) = \frac{\partial g_{i}(h_1,h_2,h_3,h_4)}{\partial h_{j}} \Biggr|_{h_{1} = h_{2} = h_{3} = h_{4} = 1}, \quad 
\]
for each $i,j=1,2,3,4$. Namely, we have

\begin{widetext}
\begin{align}
\mathbf{J} = \left[
    \begin{matrix}
        T_{f}\frac{\left<d_{fs}^{2} - d_{fs}\right>}{\langle d_{fs} \rangle}   &   T_{f}\frac{\left<d_{fs}n_{ft}\right>}{\langle d_{fs} \rangle}    &  T_{f}\frac{\left<d_{fs}d_{ws}\right>}{\langle d_{fs} \rangle} & T_{f}\frac{\left<\dfs\nwt\right>}{\langle \dfs \rangle} \\
        2T_{f}(1 + T_{f} - T_{f}^2)\frac{\left<d_{fs}n_{ft}\right>}{\langle n_{ft} \rangle}   &   2T_{f}(1 + T_{f} - T_{f}^2)\frac{\left<n_{ft}^{2} - n_{ft}\right>}{\langle n_{ft} \rangle}    &   2T_{f}(1 + T_{f} - T_{f}^2)\frac{\left<n_{ft}d_{ws}\right>}{\langle n_{ft} \rangle} & 2T_{f}(1 + T_{f} - T_{f}^2)\frac{\left<\nft\nwt\right>}{\langle \nft \rangle}\\
        T_{w}\frac{\left<d_{ws}d_{fs}\right>}{\langle d_{ws} \rangle}   &   T_{w}\frac{\left<d_{ws}n_{ft}\right>}{\langle d_{ws} \rangle}   &   T_{w}\frac{\left<d_{ws}^{2} - d_{ws}\right>}{\langle d_{ws} \rangle} & T_{w}\frac{\left<d_{ws}\nwt\right>}{\langle d_{ws} \rangle}  \\
        2T_{w}(1 + T_{w} - T_{w}^2)\frac{\left<\nwt\dfs\right>}{\langle \nwt \rangle}   &   2T_{w}(1 + T_{w} - T_{w}^2)\frac{\left<\nwt\nft\right>}{\langle \nwt \rangle}   &   2T_{w}(1 + T_{w} - T_{w}^2)\frac{\left<\nwt\dws\right>}{\langle \nwt \rangle} & 2T_{w}(1 + T_{w} - T_{w}^2)\frac{\left<\nwt^{2} - \nwt\right>}{\langle \nwt \rangle}\\
    \end{matrix}
    \right].
    \label{eq:jacobianLinearization}
\end{align}
\end{widetext}

Now, if the largest eigenvalue in absolute value of the Jacobian matrix $\mathbf{J}$, denoted by $\sigma(\mathbf{J})$, is less than or equal to one, then the trivial solution mentioned above is an \textit{attractor}, whence all informed components have finite size as understood from the conservation of probability; i.e., from $H(1)=1$. However, if
$\sigma(\mathbf{J}) > 1$, then the trivial solution will not be stable and another solution with $h_1,h_2,h_3,h_4<1$ will exist. This then will lead to having $H(1)<1$ meaning that information epidemics take place with probability $1-H(1)>0$ and reach an expected fraction $1-H(1)$ of the whole population, where $H(1)$ is computed from 
(\ref{eq:H_with_new_notation}).

Collecting, the threshold of information epidemics is given by $\sigma(\mathbf{J}) = 1$, where $\sigma(\mathbf{J})$ is the {\em spectral radius} of the Jacobian matrix given at (\ref{eq:jacobianLinearization}). Also, the mean epidemic size (i.e., the fractional size of the giant component of the percolated network $\tilde{\mathbb{H}}$) can be computed by first finding the pointwise smallest solution of the recursion (\ref{eq:recursion_with_new_notation}), and then reporting the result into (\ref{eq:H_with_new_notation}) to get $H(1)$. As discussed before, the mean size of epidemics is given by $1-H(1)$.

\subsection{The Relationship between Our Analysis and Some Previous Studies}
\label{sec:relationshipBetweenPreviousStudy}
Our results generalize some of the existing work in the literature; e.g., see \cite{MN09, JM09, OY13}. First,
by letting $h_{fs}(x)=h_{ft}(x)=1$ in (\ref{func:hfsx})-(\ref{func:hwtx}), we ensure that $\bf$ is an {\em empty} graph, so that our system model is equivalent to the single clustered network
considered in \cite{MN09, JM09}.
Similarly, if we set $h_{ft}(x) = h_{wt}(x) = 1$ then neither $\bf$ nor $\bw$ will have \textit{triangle edges}, rendering our system to be equivalent to the non-clustered multi-layer network studied in
\cite{OY13}. A careful inspection of our results will reveal that in both special cases, our results recover the finding of 
\cite{MN09, JM09, OY13}.

\section{Numerical Results and Discussion}
\label{sec:numericalResults}

This section is devoted to presenting numerical results with regard to information propagation in clustered multi-layer networks in specific settings with given degree distributions. In what follows, we first consider a simple case where both constituent networks in our model has doubly-Poisson degree distributions $p_{st}$, while in Section \ref{subsec:power_law_numerics} we consider the more realistic case where $p_{st}$ is a power-law degree distribution with exponential cut-off. Section \ref{subsec:clustering} and Section \ref{subsec:alpha} are devoted
to understanding the impact of clustering and of the parameter $\alpha$ on the dynamics of information propagation, respectively.

\subsection{Networks with Doubly Poisson Distributions}

Consider the case where both $p_{st}^{f}$ and $p_{st}^{w}$ are doubly Poisson; i.e., 
the number of \textit{single edges} and \textit{triangles} in both networks are independent and they all follow a Poisson distribution. Namely, we set
\begin{align}
p_{st}^{f} = e^{-\mu_{f,1}}\frac{(\mu_{f,1})^{s}}{s!}e^{-\mu_{f,2}}\frac{(\mu_{f,2})^{t}}{t!}, \mbox{ \ s, t = 1, 2, \dots },
\label{func:poiPfst}
\end{align}
\par
and
\begin{align}
p_{st}^{w} = e^{-\mu_{w,1}}\frac{(\mu_{w,1})^{s}}{s!}e^{-\mu_{w,2}}\frac{(\mu_{w,2})^{t}}{t!}, \mbox{ \ s, t = 1, 2, \dots },
\label{func:poiPwst}
\end{align}
where $s$ and $t$ are the number of \textit{single edges} and \textit{triangles} in the corresponding networks while $\mu_{f,1}$ and $\mu_{f,2}$ (resp.~$\mu_{w,1}$ and $\mu_{w,2}$) are the mean number of them respectively in $\bf$ (resp.~in $\bw$).

Under this setting, the mean epidemic size as well as the epidemic threshold can be computed from the
analytical results 
presented in Section \ref{sec:fractionOfTheGiantConnectedComponent}. 
To check the validity of our analysis for finite-sized networks, we have also conducted an extensive numerical study. In particular, we consider $n = 5 \times 10^{5}$ nodes in the population and three different values $\alpha = 0.1, 0.5, 0.9$ for 
the size of network $\bf$. We let 
$\mu_{f,1}= \mu_{f,2} = \lambda_{fs} = \lambda_{ft} = 0.5$ and similarly
$\mu_{w,1}= \mu_{w,2} = \lambda_{ws} = \lambda_{wt} = 0.5$. For various information transmissibility parameters $T_w=T_f$
we generate 100 independent realizations of the multi-layer network
$\mathbb{H}$ and compute the size of the largest connected component (of the percolated network $\tilde{\mathbb{H}}$) 
in each case. The results are then averaged over 100 experiments to obtain the {\em empirical} size of information epidemics. 

\begin{figure}[b]
	\centering
    \includegraphics[width=0.45\textwidth]{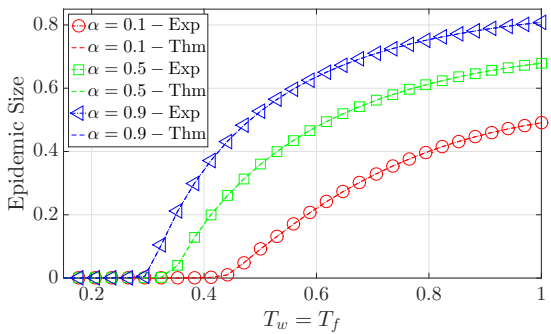}
    \caption{Simulation for doubly Poisson degree distributions.}
	\label{fig:expPoissonSim}
\end{figure}
The results are depicted in Figure \ref{fig:expPoissonSim}, where
the curves stand for the theoretical results obtained from our discussion in Section \ref{sec:fractionOfTheGiantConnectedComponent}, while the markers stand for the empirical results obtained from simulation experiments. We see that there is a perfect agreement between the analytical and experimental results confirming the validity of our results even when $n$ is finite.
We also see that as $\alpha$ increases, the critical threshold is reduced and the epidemics size is enlarged. This is
an intuitive consequence given that the network becomes {\em denser} with increasing $\alpha$. A more detailed discussion on the impact of the parameter $\alpha$ on the characteristics of information propagation in a multi-layer network is provided in Section \ref{subsec:alpha} below. 

\subsection{Networks with Power-law Degree Distributions}
\label{subsec:power_law_numerics}
Many real-world networks including the Internet (at the level of autonomous systems), the phone call network, the e-mail network, and the web link network are shown to exhibit power law degree distributions with exponential cut-off \cite{AC2009}.
To gain more insight about our results for more realistic network models, we next consider the case where both $\bf$ and $\bw$ have power-law degree distribution with exponential cut-off.
Namely, we have
\begin{align}
p_{st}^{f} = 
\begin{cases}
    0, &\mbox{s = 0 \text{or} t = 0},\\
    \frac{s^{-\gamma_{f,1}}}{\text{Li}_{\gamma_{f,1}}(e^{-1/\Gamma_{f,1}})}
    \frac{t^{-\gamma_{f,2}}}
    {\text{Li}_{\gamma_{f,2}}(e^{-1/\Gamma_{f,2}})}, &\mbox{s, t = 1, 2, \dots }
\end{cases}
,
\label{func:pfst}
\end{align}
and
\begin{align}
p_{st}^{w} = 
\begin{cases}
    0, &\mbox{s = 0 \text{or} t = 0},\\
    \frac{s^{-\gamma_{w,1}}}{\text{Li}_{\gamma_{w,1}}(e^{-1/\Gamma_{w,1}})}
    \frac{t^{-\gamma_{w,2}}}{\text{Li}_{\gamma_{w,2}}(e^{-1/\Gamma_{w,2}})},
    &\mbox{s, t = 1, 2, \dots }
\end{cases}
,
\label{func:pwst}
\end{align}
where $\text{Li}_{m}(z) = \sum_{k = 1}^{\infty}\frac{z^{k}}{k^{m}}$ is the $m^{th}$ polylogarithm of $z$.

In order to compute an analytical expression for
the size of information epidemics we proceed similarly with the case
of doubly Poisson distributions and use our results presented in Section \ref{sec:fractionOfTheGiantConnectedComponent}. 

\begin{figure}[!t]
	\centering
    \includegraphics[width=0.5\textwidth]{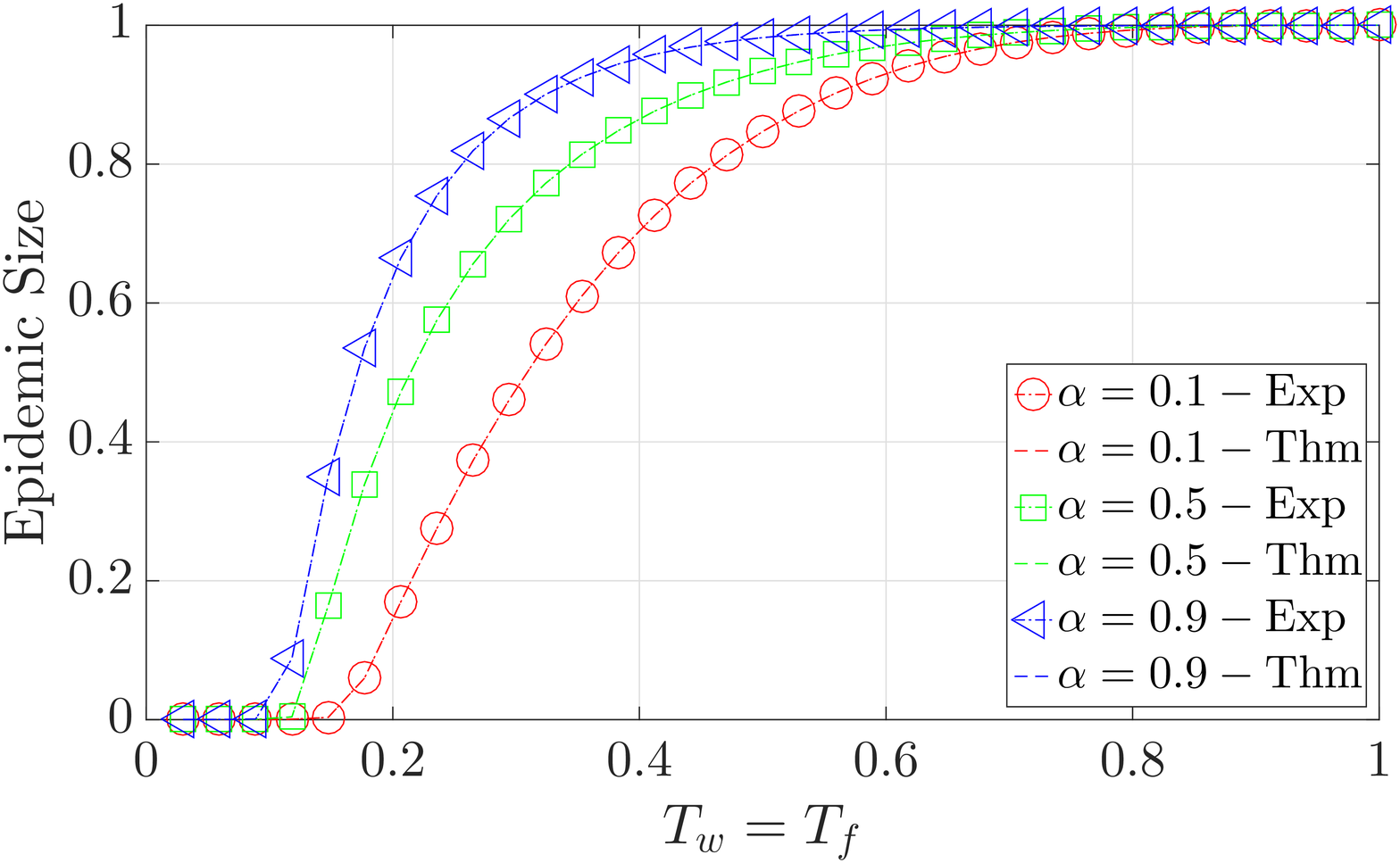}
    \caption{Simulation for Power-law degree distributions.}
	\label{fig:expPowerSim}
\end{figure}

For computer simulations, we again set $n = 2 \times 10^{5}$,
and use $\alpha = 0.1, 0.5, 0.9$ as three sample sizes for the network $\bf$. The corresponding degree distributions 
are given by (\ref{func:pfst}) and (\ref{func:pwst})
with
$\gamma_{f,1} = \gamma_{f,2} = \gamma_{w,1} = \gamma_{w,2} = 2.5$, and $\Gamma_{f,1} = \Gamma_{f,2} = \Gamma_{w,1} = \Gamma_{w,2} = 10$. With $T_{f}=T_{w}$ ranging from zero to one, we compute the empirical size of the information epidemics again via averaging over 100 independent experiments. The results are demonstrated
in Figure \ref{fig:expPowerSim} where curves are obtained analytically using our discussion
in Section \ref{sec:fractionOfTheGiantConnectedComponent}, and markers represent numerical results.
We again see a perfect agreement between our analysis and numerical results. 

%

\subsection{How does Clustering Affect the Threshold and Size of Information Epidemics?}
\label{subsec:clustering}

An important goal of this work is to understand how clustering affects the dynamics of information propagation in multi-layer networks. Given the complexity of the model adopted here this can be studied in several different ways; e.g. with controlling the clustering coefficient of only one of the networks $\bf$ or $\bw$, or by adjusting both networks' clustering simultaneously. Also, the way we adjust the clustering coefficient of a given network can have a significant impact on the conclusions obtained given that such changes might also impact the degree-degree correlations (e.g., assortativity) in the network\footnote{The assortativity coefficient is the Pearson correlation coefficient of degree between pairs of linked nodes, and the detail of the computation can be found in \cite{MN02a}.}. The situation becomes even more involved as one realizes that the choice of the parameter $\alpha$  changes the assortativity of the network as well.




With these in mind, we consider doubly Poisson distributions in the remainder of this discussion. We first consider a scenario where 
one or both of the constituent networks in the system is changed from a non-clustered network to a clustered network. More precisely, we compare the following three cases
\begin{itemize}
\item Both networks are non-clustered (NN)
\item Network $\bw$ is clustered but network $\bf$ is non-clustered (NC)
\item Both networks are clustered (CC)
\end{itemize}
Here, the clustered networks are generated as discussed in Section
\ref{sec:networkClustering} following the approach of Miller \cite{JM09} and Newman \cite{MN09}, say with doubly Poisson degree distribution $p_{st}$ with parameter $\lambda_s$ for single edges and $\lambda_t$ for triangle edges. To ensure a fair comparison, we generate non-clustered networks with  the same total degree distribution {\em and}  degree-degree correlations. To this end, we generate the non-clustered networks using the multiplex (i.e., colored) version of the configuration model \cite{BS02}. Namely, each node gets $\textrm{Poi}(\lambda_s)$ stubs of color {\em blue}
and $2 \times \textrm{Poi}(\lambda_t)$ stubs of color {\em red}, and then stubs of the same color are randomly matched to form edges. The standard configuration model where colors are ignored would lead to the same degree distribution, but would fail in capturing the positive degree correlations inherent in the random clustered networks proposed by Miller \cite{JM09} and Newman \cite{MN09}.





\begin{figure}[!t]
	\centering
    \includegraphics[width=0.45\textwidth]{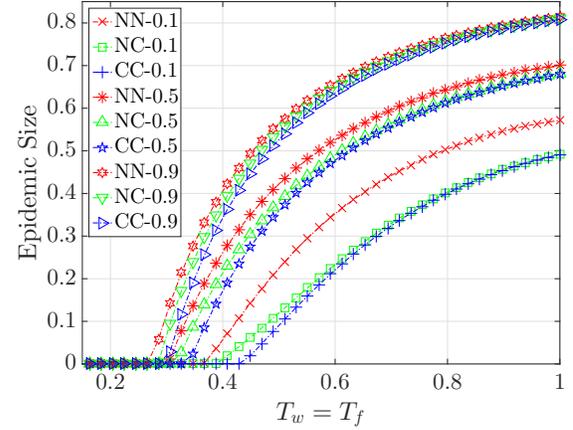}
    \caption{Comparison of the size of information epidemics between Non-clustered and Non-clustered networks (NN), Clustered and Non-clustered networks (NC), and Clustered and Clustered networks (CC). Plots are obtained from our analytical results.
    The value following the model abbreviation indicates the amount of overlapping between two networks. For example, NC-0.5 means that $\alpha=0.5$.
}
	\label{fig:expNNNCCC}
\end{figure}

The results are depicted in Figure \ref{fig:expNNNCCC}, where 
we compare the relative size of the epidemics as $T_w=T_f$ varies from zero to one, and $\alpha$ is taken to be $0.1$, $0.5$, or $0.9$.
The resulting \textit{global} and \textit{local} clustering coefficients, and the assortativity values of network $\mathbb{H}$ can be found in Table \ref{table:clusteringCoefficientsSameDistributation}. For each $\alpha$ value, we see that clustering increases as we go from $NN$ to $NC$ to $CC$, while assortativity stays the same. 
Our main conclusion from  Figure \ref{fig:expNNNCCC} is that for a given $\alpha$, the curve for the NN is always above that of NC, which in turn is always above that of CC.
That is, the critical threshold of information epidemics increases while the final epidemic size decreases as we move from $NN$ to $NC$ to $CC$, i.e., as the clustering coefficient in the whole system increases. Therefore, we conclude that the high level of clustering not only makes it more difficult for information to reach a significant fraction of the population, but it also reduces the mean epidemic size at any level of information transmissibility.

The inhibitive effect of clustering on epidemics has been observed in the single network case as well \cite{JM09}, and is often attributed to the fact that the edges used for completing wedges to triangles is redundant for the purposes of information propagation; a wedge is defined as a connected triple that is not a triangle. This is particularly evident when $T_{f}=T_{w}=1$, and the size of the epidemics is equal to the giant component size in $\mathbb{H}$. It is clear that adding an extra edge to this graph that transforms a wedge into a triangle has no effect on its giant component; in contrast it may be possible to increase the giant component size  by adding this extra edge somewhere else in the network. Therefore, as long as the degree distributions and degree-degree correlations are fixed, random networks with low clustering will tend to have a larger epidemic size and a lower epidemic threshold.



\begin{table*}[!t]
    \centering
    \begin{tabular}{l|c|c|c|c|c|c|c|c|c|c}
                    &   \multicolumn{3}{c|}{Non-clustering and Non-clustering} &  \multicolumn{3}{c|}{Non-clustering and Clustering}  &   \multicolumn{3}{c|}{Clustering and Clustering}\\
        \hline
        $\alpha$    &   Global Coeff.    &    Local   Coeff.    &   Assortativy&   Global Coeff.  &   Local Coeff.  &   Assortativy  &   Global Coeff.  &   Local Coeff.    &   Assortativy  \\
        \hline
        0.1         & 0   &   0   & 0.106   &   0.25    & 0.48  &   0.106   &   0.27    &   0.51    &   0.106\\
        0.5         & 0   &   0   & 0.071   &   0.14    & 0.31  &   0.071   &   0.21    &   0.42    &   0.071\\
        0.9         & 0   &   0   & 0.044   &   0.01    & 0.19  &   0.044   &   0.19    &   0.35    &   0.044
    \end{tabular}
    \caption{Statistics of network $\mathbb{H}$ under the setting of Figure \ref{fig:expNNNCCC}.
    }
    \label{table:clusteringCoefficientsSameDistributation}
\end{table*}

In order to understand the effect of clustering better, we next consider a different setting where we control the level of clustering in network 
$\bw$ while keeping its mean total degree fixed. More precisely,
we use Poisson distributions for the number of single and triangle edges in both networks with parameters given in Table \ref{table:parametersDoublyPoissionDistribution}. Put differently, network $\bf$ has a fixed clustering coefficient while 
with $c \in [0, 4]$ the clustering of $\bw$ varies between the two extremes: i) when $c = 4$, $\bw$ will have no {\em single}-edges and consist only of \textit{triangles} resulting with a clustering coefficient close to one; and ii)
with $c = 0$, there will be no triangles in $\bw$ and hence its clustering coefficient will be close to zero. Thus, with increasing $c$, the clustering coefficient of $\bw$ increases, which in turn increases clustering in the multilayer network $\bh$; see Table \ref{table:statisticsCompareCC} for specific clustering coefficients corresponding to several $c$ values considered. We remark that by the choice given in Table \ref{table:parametersDoublyPoissionDistribution}, the degree distribution (single edges plus triangle edges) of $\bw$ is given by 
\[
2 \textrm{Poi}\left(\frac{4 - c}{2}\lambda\right) + 2 \textrm{Poi}\left(\frac{c}{2}\lambda\right).
\]
This ensures that as $c$ varies both the mean and the variance of the degree distribution remains constant, allowing us to focus only on the effect of clustering; for instance, using $\textrm{Poi}((4-c)\lambda)$ rather than  
$2 \textrm{Poi}\left(\frac{4 - c}{2}\lambda\right)$ would change the variance of the distribution and hence the threshold for information epidemics (viz.~(\ref{eq:criticalPointSingleLayerNetwork})). As seen from Table \ref{table:parametersDoublyPoissionDistribution}, the assortativity of the network also remains constant with varying $c$.

\begin{table}[!ht]
    \centering
    \begin{tabular}{l|c|c}
                                            &   Network $\bf$   &   Network $\mathbb{W}$\\
        \hline
        Distribution of \textit{single}-edges   &   Poi($2\lambda_{\bf}$)  &   2  Poi$(\frac{4 - c}{2}\lambda_{\bw})$              \\
        & & \\
       Distribution of \textit{triangles} &   Poi($\lambda_{\bf}$)   &   Poi$\left(\frac{c}{2}\lambda_{\bw}\right)$
    \end{tabular}
    \caption{Parameters of the doubly Poisson distribution.
    In Figure \ref{fig:compareCC} we set $\lambda_{\bf} = \lambda_{\bw} = 0.5$.
    We use $\lambda_{\bf} = 0.36$ and $\lambda_{\bw} = 0.5$ for Figure \ref{fig:compareGcDifferentLambdaFW}.
    }
    \label{table:parametersDoublyPoissionDistribution}
\end{table}

\begin{figure}[!ht]
    \centering
    \includegraphics[width=0.45\textwidth]{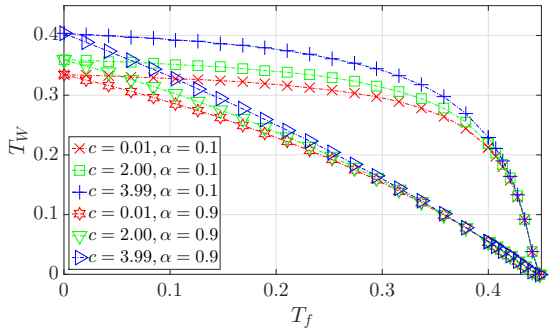}
        \caption{
        Comparison of the epidemic {\em boundary} under several cases; the north and east of each curve specifies
        the region of $(T_f,T_w)$ values for which epidemics are possible, while the south and west part of each
        curve stands for the region where epidemics can {\em not} take place.
        Resulting statistics for clustering and assortativity is given in Table \ref{table:statisticsCompareCC}.
        }
    \label{fig:compareCC}
\end{figure}

With these in mind, we first demonstrate in Figure \ref{fig:compareCC} the boundary of the $T_f-T_w$ plane that identifies the threshold of information epidemics. Put differently, for each parameter pair $(c,\alpha)$, the curves in Figure \ref{fig:compareCC} separates the region where information epidemics can take place (north and east of the curves) from the region where they can not (south-west of the curves). We see that with the same $T_{f}$, clustering increases the minimum  $T_{w}$ that is needed for information epidemics to be possible. In other words, we see again that clustering increases the 
threshold of epidemics.

\begin{figure}[h]
    \centering
    \includegraphics[width=0.45\textwidth]{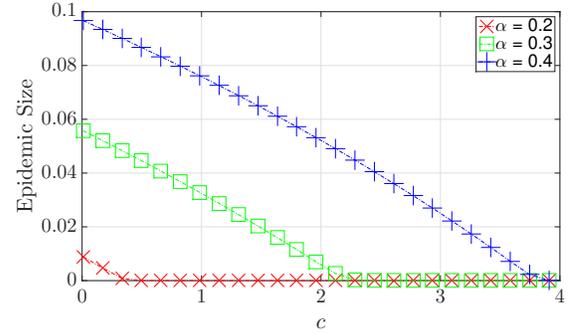}
    \caption{Illustration of how clustering affects the size of epidemics when $T_f = T_w = 0.3$.
    }
\label{fig:compareGcDifferentLambdaFW}
\end{figure}

Next, we look at the effect of clustering on the relative final size of information epidemics for specific percolation (i.e., transmissibility) probabilities.
From Figure \ref{fig:compareGcDifferentLambdaFW}, we see that the size of giant component decreases as the clustering coefficient increases, again confirming that 
high clustering reduces the epidemic size.


\begin{table}[!h]
    \centering
    \begin{tabular}{c|c|c|r|r|}
        \multirow{2}{*}{$\alpha$}   &   \multirow{2}{*}{$c$}&   \multirow{2}{*}{assortativity}&  \multicolumn{2}{c|}{Clust. Coefficients}\\
                                    &                       &                               &  Global  &   Local\\
       \hline
       \multirow{3}{*}{0.1}         &   0.01                &   0.010&  0.005   &   0.006\\
                                    &   2.00                &   0.010&  0.095   &   0.230\\
                                    &   3.99                &   0.010&  0.185   &   0.453\\
       \hline
       \multirow{3}{*}{0.9}         &   0.01                &   0.009&  0.023&  0.044\\
                                    &   2.00                &   0.009&  0.075&  0.152\\
                                    &   3.99                &   0.009&  0.126&  0.260\\
    \end{tabular}
    \caption{Statistics corresponding to the network $\mathbb{H}$ in the setting of Figure \ref{fig:compareCC}.}
    \label{table:statisticsCompareCC}
\end{table}

\subsection{How does $\alpha$ Affect the Information Propagation Dynamics?}
\label{subsec:alpha}

We now shift our focus to understanding the impact of the parameter $\alpha$, which controls the relative size of network $\bf$ to network $\bw$, on the information propagation dynamics. From a practical perspective, this will help understand the role of the size of an online social network, say Facebook, on propagating the information. As we shall demonstrate soon, this parameter's impact on the overall network topology goes beyond a change in degree distribution, and thus its effect on epidemic threshold and epidemic size are highly non-trivial. 


In order to focus only on the impact of $\alpha$, we consider non-clustered networks throughout this section; however, a similar discussion would hold for clustered networks as well. 
Let $\bw$ and $\bf$ be random networks with given degree distributions, and let $\mathbb{H} = \bw \coprod \bf$ be their disjoint union; i.e., network $\mathbb{H}$ is a colored degree-driven random graph introduced in \cite{BS02}. As before $\bw$ is defined on $n$
vertices, each of which belongs to vertex set of $\bf$ independently with probability $\alpha$. For simplicity, we assume that network $\bw$ 
has Poisson degree distribution with parameter $\lambda_{\text{red}}$,
while network $\bf$ has degree distribution $\text{Poi}(\lambda_{\text{blue}})$. Under this setting, each of the $n$ nodes in graph $\mathbb{H}$ will have a colored degree distribution given by
\begin{align}
\label{degree_dist_colored_osy}
\begin{cases}
    p_{k}^{\text{red}} = e^{-\lambda_{\text{red}}}\frac{\lambda_{\text{red}}^{k}}{k!}, \quad \mbox{ \ k = 0, \dots },    \\
    p_{k}^{\text{blue}} = \alpha e^{-\lambda_{\text{blue}}}\frac{\lambda_{\text{blue}}^{k}}{k!} + (1 - \alpha)\mathbf{1}[k = 0], \quad \mbox{ \ k = 0, \dots }.
\end{cases}
\end{align}
where blue edges represent links in $\bf$ and red edges represent links in $\bw$. The multiplex network (MN) $\mathbb{H}$ is then generated by the {\em colored} configuration model where only stubs of the same color are connected together to form an edge. 

To check the impact of $\alpha$ on the size of information epidemics in a fair way, we keep the value of $\alpha \lambda_{blue}$ fixed throughout the experiments.
This ensures that the mean number of blue edges in $\mathbb{H}$ remains constant as $\alpha$ varies. Put differently, this setting allows us to 
compare the impact of a {\em small} but densely connected social network with a {\em large} but loosely connected one in facilitating the propagation of information. Below, we will argue why an adjustment on $\alpha$ changes not only the degree distribution but also the degree-degree correlations in the network. To make this point clearer, we also include in our comparison the simplex network (SN) case which ignores the colors of the edges and generates $\mathbb{H}$ via the standard configuration model with degree distribution $p_k =  p_{k}^{\text{red}} \oplus p_{k}^{\text{blue}}$; here $\oplus$ denotes the convolution operator.

\begin{figure}[!t]
    \centering
    \includegraphics[width=0.45\textwidth]{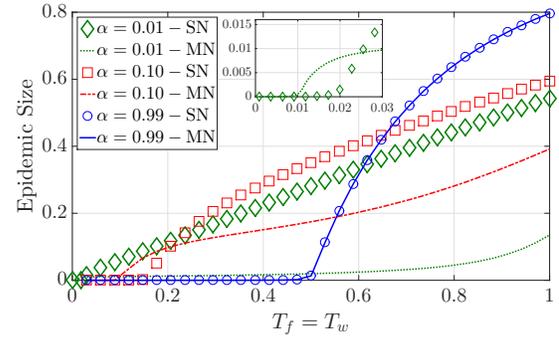}
    \caption{Illustration of the effect of $\alpha$.
    SN is the abbreviation for Simplex Network where colors of the edges are ignored, while MN indicates the Multiplex Network case where only stubs of the same color are connected together. (Inset) The plots for $\alpha=0.01$ are shown at a higher resolution near the phase transition point.
    }
\label{fig:compareAlpha}
\end{figure}

The results comparing the final epidemic sizes for three specific $\alpha$ values are given in 
Figure \ref{fig:compareAlpha}. These plots are obtained via computer simulations with $n=5 \times 10^5$, $\lambda_{\text{red}}=1$, $\alpha \lambda_{\text{blue}}=1$, and $T_w=T_f$ is varied from zero to one; each data point corresponds to an average over 100 independent runs.
We list the resulting assortativity values for each case in Table \ref{table:assortativyOfSingelLayerNetworkAndMultilayerNetwork}. As expected, the simplex case that corresponds to the standard configuration model has uncorrelated degrees and thus the resulting assortativity is zero. However, we realize that aside from changing the degree distribution in the network, the relative size of $\bf$ also has a significant impact on the degree-degree correlations in the multiplex case. This impact, namely the positive correlations observed between the degrees of neighbors, is particularly pronounced in the case where $\alpha$ is small; e.g., assortativity is $0.96$ when $\alpha = 0.01$. This can be attributed to the fact that when $\alpha$ is close to zero, a very small fraction of nodes receive a large number of blue edges (since $\alpha \lambda_{\text{blue}}$ is fixed) and these extra edges can only be used to connect with other nodes that also have extra edges; as before red
edges are assigned to every node.  As a result, the network $\mathbb{H}$ exhibits a very densely connected (community-like) subgraph on the vertices that participate in $\bf$, and this leads to highly positive degree-degree correlations given that the nodes in $\bf$ have significantly larger (in the statistical sense) degrees than nodes that 
are not in $\bf$.

\begin{table}[!h]
    \centering
    \begin{tabular}{c|c|c|c|c|c|}
        \multicolumn{2}{c|}{$\alpha = 0.01$} &   \multicolumn{2}{c|}{$\alpha = 0.10$} &   \multicolumn{2}{c|}{$\alpha = 0.99$} \\
        \hline
        SN      &   MN      &   SN      &   MN      &   SN      &   MN  \\
        \hline
        0.00    &   0.96    &   0.00    &   0.65    &   0.00    &   0.00   
    \end{tabular}
    \caption{Comparison of the assortativity values observed in the setting of Figure \ref{fig:compareAlpha} for to the Simplex Network (SN) and the Multiplex Network (MN) case for different $\alpha$. As expected, for the simplex network case the degrees of the nodes are uncorrelated and assortativity is thus zero. The multiplex case exhibits assortative mixing, with the correlations getting more significant with decreasing $\alpha$.}
    \label{table:assortativyOfSingelLayerNetworkAndMultilayerNetwork}
\end{table}

There are a number of interesting conclusions we can derive from Figure \ref{fig:compareAlpha}. First, by comparing the simplex and multiplex cases with each other for each $\alpha$ value (i.e., by comparing the line and the marker that are of the same color in Figure \ref{fig:compareAlpha}) we see that multiplex networks have a smaller epidemic threshold as well as a smaller epidemic size as compared to the corresponding simplex network for small $\alpha$ values; on the other hand for $\alpha \simeq 1$, the differences are negligible. This observation is in line with the assortativity values seen in Table \ref{table:assortativyOfSingelLayerNetworkAndMultilayerNetwork} noting the fact that assortativity is known \cite{MN02,JM09} to reduce the critical threshold and the size of epidemics. 

Second, we focus on the impact of $\alpha$ on the threshold and size of epidemics by comparing the three lines in Figure \ref{fig:compareAlpha} that correspond to the multiplex case with $\alpha=0.01$, $\alpha=0.1$ and $\alpha=0.99$, respectively. We observe that as $\alpha$ gets larger, the epidemic threshold increases and so does the final epidemic size. What this means is that, it will be more {\em difficult} to trigger an information epidemic when the physical network is augmented with a large online social network that is loosely connected, as compared to the case when the online social network is small but densely connected. However, when information transmissibility is high in both networks, final epidemic size is going to be larger in the case of a large but loosely connected online social network as compared to the case of a small but densely connected one. Combining, we conclude that information with low transmissibility spreads more effectively with a small but densely connected social network, whereas highly transmissible information will reach more people with the help of a large but loosely connected social network; here the basis of comparison is again the total number edges in the overlay network.

It is important to remark that the differences observed between the three lines for the multiplex case may not be solely attributed to the changes in the assortativity levels. This is because when we adjust $\alpha$, the degree distribution of the network changes as well. For example, it is easy to see that as $\alpha$ increases the 
variance of the degree distribution tends to be lower, which is known \cite{MM95} to increase the epidemic threshold; i.e., it has a 
similar impact on the epidemic threshold with
reducing the assortativity. 
To better understand the impact of $\alpha$ on the degree distribution and hence on the information propagation
dynamics, we now compare the three simplex network cases in Figure \ref{fig:compareAlpha}, i.e., we compare the data points shown with markers. This time as well, we see that as $\alpha$ gets larger the epidemic threshold and the final epidemic size gets larger, although the differences observed are less significant as compared to the multiplex case discussed above. Intuitively speaking, this would be expected since none of these three cases exhibit assortativity and the observed impact is only due to the change in the degree distribution. 
The observed impact of $\alpha$ on the degree distribution and on the epidemic threshold can in fact be quantified. First, we recall that for a single layer network, the critical threshold for epidemics is given by \cite{MM95,MN02}
\begin{align}
   T \frac{\bbe[d_{i}(d_{i} - 1)]}{\bbe[d_{i}]} > 1,
         \label{eq:criticalPointSingleLayerNetwork}
 \end{align}
where $d_{i}$ is the degree of an arbitrary node $i$. With the choice of the degree distribution given in
(\ref{degree_dist_colored_osy}) and $T_f=T_w=T$, it is easy to see that this condition reduces to
\[
\alpha \lambda_{\text{blue}} + \lambda_{\text{red}}+ \frac{\alpha \lambda_{\text{blue}}^2-\alpha^2 \lambda_{\text{blue}}^2}{\alpha \lambda_{\text{blue}}+\lambda_{\text{red}}}>\frac{1}{T},
\]
or, equivalently to
\[
T > \frac{2}{3+1/\alpha}
\]
with our choices of $\alpha \lambda_{\text{blue}}=\lambda_{\text{red}}=1$. This finding quantifies how the critical threshold should increase with $\alpha$, and it is in perfect agreement with the curves for the simplex case shown in Figure \ref{fig:compareAlpha} as expected. 




\section{Conclusion}
\label{sec:conclusion}
We analyze the  propagation of information in clustered multilayer networks, where the vertex set of one network is a subset of the vertex set of the other. We solve analytically for the threshold, probability, and mean size of information epidemics, and confirm our findings via extensive computer simulations. We show from various angles that clustering increases the epidemic threshold and decreases the final epidemic size in multi-layer networks. We also demonstrate how the overlap between the constituent networks affects the information propagation dynamics, particularly through impacting the degree-degree correlations. For instance, we show that an online social network $\bf$ that is small in size but large in mean connectivity is more effective in facilitating the propagation of information as compared to a large social network with smaller mean connectivity, with the total number of edges fixed in both cases.

Our general framework contains non-clustered multi-layer networks and single-layer clustered networks as special cases. In addition, given that information propagation problem is studied via bond percolation over a multi-layer network, our work can also be useful in the context of robustness against {\em random attacks} -- Assume that our system consists of two conjoint networks $\bf$ and $\bw$ and an adversary attacks edges in both networks randomly with probabilities $T_f$ and $T_w$, respectively with the aim of disconnecting the whole system. Then, the size and existence of the giant component after edge failures would be natural metrics for the robustness of this system against random attacks. To that end, we believe our results (e.g., Figure \ref{fig:compareCC}) would be useful in understanding the impact of clustering on the robustness of multi-layer networks.

There are many open problems one might consider for future work. For instance, the impact of assortativity is not fully understood on the propagation of information over multi-layer networks. Another interesting direction would be to consider networks that exhibit clustering not only through triangles, but also through larger cliques. Extending some of the ideas presented here to the case of influence propagation (e.g., complex contagions) would also be interesting.



\section*{Acknowledgment}
This research was supported in part by National Science Foundation through grant CCF \#1422165, and in part by the Department of Electrical and Computer Engineering at Carnegie Mellon University.

\bibliographystyle{ieeetr}
\bibliography{sdp}

\begin{IEEEbiographynophoto}{Yong Zhuang}
(S'12) received the B.S. degree at Zhejiang University, Hangzhou (China) in 2012, and the M.S. degree at National Taiwan University in 2014.
He is now a PhD student at the Department of Electrical and Computer Engineering at Carnegie Mellon University.
His research interests are in network science, random graphs, social and information networks.
\end{IEEEbiographynophoto}

\begin{IEEEbiographynophoto}{Osman Ya\u{g}an}
(S'07-M'12) received the B.S. degree in Electrical and Electronics Engineering from Middle East Technical University, Ankara (Turkey) in 2007, and the Ph.D. degree in Electrical and Computer Engineering from University of Maryland, College Park, MD in 2011.
He was a visiting Postdoctoral Scholar at Arizona State University during Fall 2011 and then a Postdoctoral Research Fellow in the Cyber Security Laboratory (CyLab) at Carnegie Mellon University until July 2013.
In August 2013, he joined the faculty of the Department of Electrical and Computer Engineering at Carnegie Mellon University as an Assistant Research Professor.
Dr. Ya\u{g}an's research interests include wireless communications, security, random graphs, social and information networks, and cyber-physical systems.
\end{IEEEbiographynophoto}






\end{document}